\documentclass[reprint,frontmatterverbose,preprintnumbers,nofootinbib,amsmath,amssymb,aps,prl,rmp,prstab,prstper,floatfix]{revtex4-2}

\usepackage[T1]{fontenc}
\usepackage{newtxtext}
\usepackage[colorlinks=true,linkcolor=blue, urlcolor=blue,citecolor=blue, anchorcolor=blue]{hyperref}     
\usepackage{physics}
\usepackage{graphicx}
\usepackage{dcolumn}
\usepackage{bm}
\usepackage[mathlines]{lineno}
\AtBeginDocument{\addtolength{\textheight}{1.0cm}}

\renewcommand{\a}{\boldsymbol{\alpha}}
\newcommand{\kk}{\!}
\begin{document}
\title{Weak Values in Postselected Quantum Trajectories}

\author{Janarbek Yuanbek} 
\thanks{These authors contributed equally to this work.} 
\author{Yuan-De Jin} 
\thanks{These authors contributed equally to this work.}
\author{Wen-Long Ma}
\email{wenlongma@semi.ac.cn}
\affiliation{State Key Laboratory of Semiconductor Physics and Chip Technologies, Institute of Semiconductors, Chinese Academy of Sciences, Beijing 100083, China}
\affiliation{Center of Materials Science and Opto-Electronic Technology, University of Chinese Academy of Sciences, Beijing 100049, China}
\date{\today}
\begin{abstract}
Weak value (WV) describes the amplified response of a meter when the measured system undergoes postselection. While conventional WV theory mainly rests on a single system–meter interaction setting, it remains unexplored whether WV can act as an effective pointer for sequential quantum measurements under postselection. Here we introduce the concept of WVs to the setting of sequential quantum measurements and quantum trajectories. We find that WV naturally emerges in postselected quantum trajectories of sequential weak measurements, by establishing a connection between the WV of the measured observable and the postselected statistics of quantum trajectories. Using sequential weak measurements on a qubit or an oscillator implemented by Ramsey interferometry with an ancilla qubit, we show that postselection shifts the collective record polarization away from the unconditioned ones, with an effective bias described by the WV. Our predictions are feasible to test in various platforms. 
\end{abstract}
\maketitle

\paragraph{Introduction.} Since its first introduction~\citep{Aharonov1988}, the concept of weak value (WV) has attracted considerable attention, both in fundamental quantum measurement theory~\citep{Dressel2014,Kedem2010,PhysRevLett.109.150407,PhysRevA.91.032116,PhysRevA.107.063706,PhysRevA.103.062206} and for weak-value amplification (WVA) ~\citep{science.1152697,PhysRevLett.105.010405,PhysRevLett.102.173601,Magana-Loaiza2014,PhysRevLett.125.080501,PhysRevLett.126.220801}. The conventional formulation considers a single system-meter interaction setting, where the WV determines the response of the meter when measuring a system observable~\cite{Dressel2014}. In particular, when
the system weakly coupled to the meter is preselected
and postselected, the average meter response, governed by the
real part of WV, can exceed the range determined by the spectrum of the measured observable, thus enabling WVA.

Beyond this conventional single measurement setting, many protocols involve sequential weak measurements (WMs) of a quantum system, with applications in probing time-ordered quantum amplitudes~\citep{PhysRevA.76.062105}, measuring incompatible observables~\citep{PhysRevLett.117.170402,Kim2018}, and high-resolution quantum sensing~\citep{Pfender2019NatCommun,Cujia2019Tracking}.
Repeated WMs with postselection at each step can exhibit WV-controlled dynamical discontinuities~\citep{394k-lj6s}.
Related direct-measurement schemes can probe virtual quantum-history amplitudes~\citep{GeorgievCohen2018} and reconstruct quantum propagators~\citep{wen2023demonstration}.
Weak values have also been studied in sequential unsharp measurements~\citep{PhysRevA.94.010103,PhysRevA.107.063706} and continuous-time measurement-output distributions~\citep{Wiseman2002Trajectories,PhysRevLett.111.160401,PhysRevA.92.012119,FranquetNazarov2017}. However, it remains largely explored how a final postselection reshapes the statistics of quantum trajectories in sequential discrete-time WMs, and whether WV can appear as an effective pointer in such postselected quantum trajectories.

\begin{figure}[t]
\centering
\includegraphics[width=\linewidth]{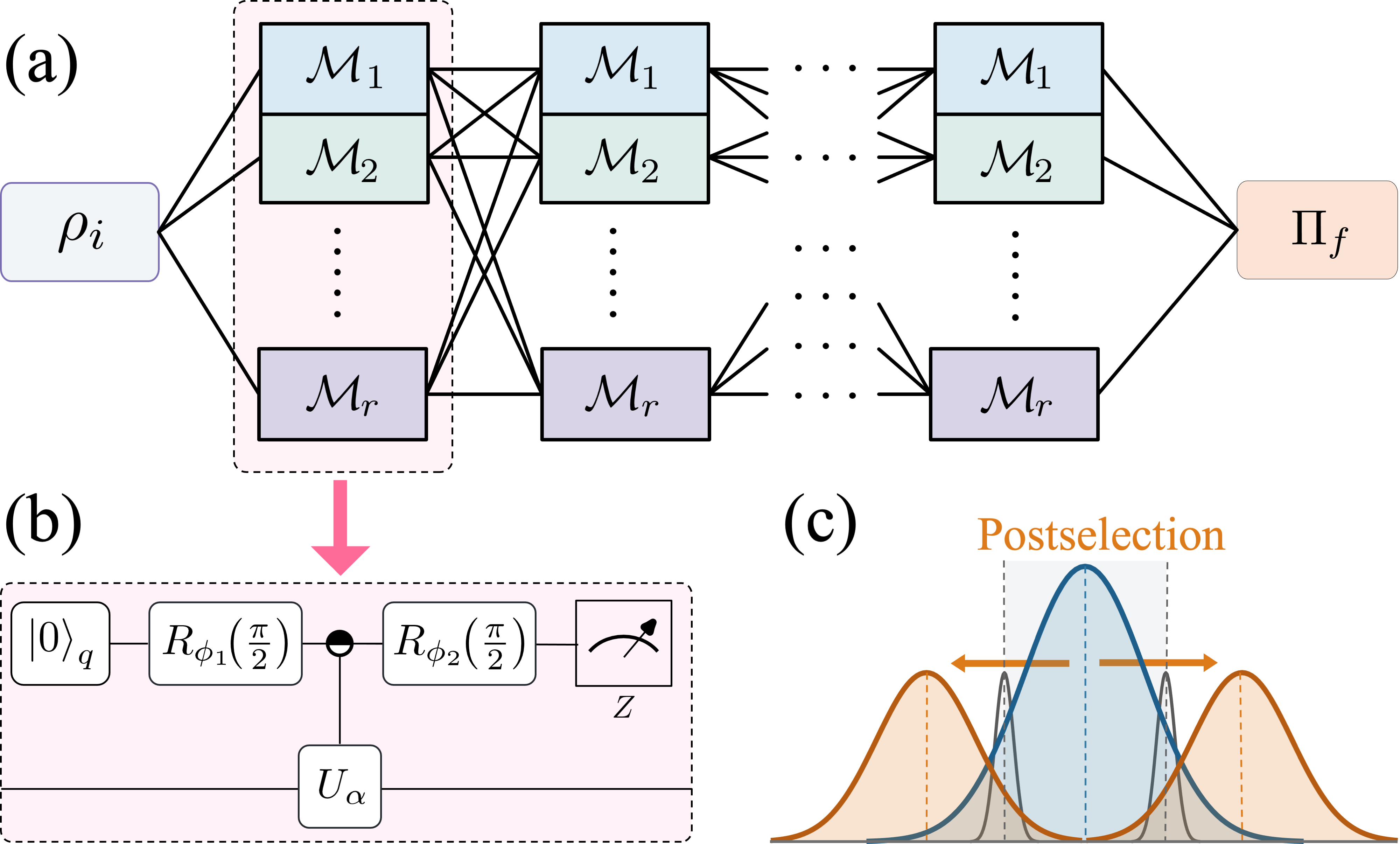}
\caption{
WVs in postselected quantum trajectories. (a) Repeated application of a quantum instrument $\{\mathcal M_\alpha\}_{\alpha=1}^{r}$ to the initial state $\rho_i$ generates quantum trajectories labeled by complete measurement records. The target is postselected by $\Pi_f$ after the final measurement. (b) Ramsey implementation of a binary weak measurement using an ancilla qubit coupled to the target. (c) Schematic distributions of the record polarization $X$. The blue curve shows the unconditioned ensemble. The gray curves, centered at the gray dashed lines, mark the conventional limits on the mean response set by the smallest and largest eigenvalues populated in $\rho_i$. Final-state postselection reweights the complete records, producing the orange conditional distributions whose means can lie beyond these limits.
\label{fig:channel} }
\end{figure}

In this paper, we show how WVs are encoded in the statistics of quantum trajectories generated by sequential WMs with final-state postselection [Fig.~\ref{fig:channel}{\color{blue}(a)}]. 
Using the quantum-instrument formalism, we relate conditional averages of measurement records to WVs of the accumulated dynamical generator. We illustrate this response for a qubit and an oscillator using Ramsey interferometry measurements (RIMs) on an ancilla qubit [Fig. \ref{fig:channel}{\color{blue}(b)}]. For identical symmetric binary WMs of a fixed observable, the collective record polarization serves as an effective pointer. Its leading conditional bias is governed by the real part of the WV of the measured observable [Fig. \ref{fig:channel}{\color{blue}(c)}], and WV appears as an effective bias of the collective measurement record. 
Our results thus characterize the collective record polarization as an effective WV pointer and quantify its finite-sequence range of validity.

\paragraph{General setting.} 
We describe each measurement step by a quantum instrument~\citep{Davies1970,Ozawa1984}, i.e., a set of completely positive (CP) maps $\{\mathcal{M}_\alpha\}_{\alpha=1}^r$ labeled by the measurement outcome $\alpha$, with $\mathcal{M}_\alpha(\rho)=M_\alpha\rho M_\alpha^\dagger$ and Kraus operators satisfying completeness condition $\sum_\alpha M_\alpha^\dagger M_\alpha=\mathbb{I}$. The instrument provides both the measurement probability and the state update rule. The measurement outcome $\alpha$ occurs with a probability $p(\alpha|\rho)=\Tr[\mathcal{M}_\alpha(\rho)]$, meanwhile steering the state to $\rho'=\mathcal{M}_\alpha (\rho) /p(\alpha|\rho)$ . Discarding the outcome yields the nonselective map $\Phi=\sum_\alpha\mathcal{M}_\alpha$, i.e., $\Phi(\rho)=\sum_\alpha M_\alpha\rho M_\alpha^\dagger$, which is CP and trace-preserving and hence a quantum channel~\citep{Wilde_2017,RevModPhys.86.1203}. Note that a quantum channel admits different quantum instruments related by unitary mixing of the Kraus operators~\citep{PhysRevA.107.012217}, while measurement statistics are determined by the specified instrument $\{M_\alpha\}$.

Repeating the instrument $m$ times produces a measurement record $\a=(\alpha_1,\ldots,\alpha_m)$, which labels a discrete-time quantum trajectory of the system [Fig.~\ref{fig:channel}{\color{blue}(a)}]. Each trajectory can be described by the CP map $\mathcal{K}_{\a}=\mathcal{M}_{\alpha_m}\cdots\mathcal{M}_{\alpha_1}$ with the Kraus operator $K_{\a}=M_{\alpha_m}\cdots M_{\alpha_1}$, and summing over all records recovers the sequential channel, $\Phi^m=\sum_{\a}\mathcal{K}_{\a}$. When incorporating the final-state post-selection,  the joint probability of obtaining the record $\a$ and then successfully postselecting the system is~\citep{Dressel2014,kfn4-1fx4,PhysRevLett.111.160401}
$
P_{\text{post}}(\boldsymbol{\alpha}) = \text{Tr} \left[ \Pi_f K_{\boldsymbol\alpha} \rho_i K_{\boldsymbol\alpha}^\dagger \right],
$
where the initial state is $\rho_i$, and the final postselection is described by $\Pi_f=\ket{\psi_f}\bra{\psi_f}$.  Its sum over all records is the success probability $P_f=\operatorname{Tr}[\Pi_f\Phi^m(\rho_i)]$, and $P_{\rm ps}(\a)=P_{\rm post}(\a)/P_f$ is the distribution of the records with successful postselection.

Here we consider the discrete-time WM, in which each step just perturbs the quantum state slightly. We quantify the measurement strength by a dimensionless parameter $\epsilon\ll1$, such that at $\epsilon=0$ every Kraus operator is proportional to the identity, for which the outcome $\alpha$ occurs with a probability $p_\alpha$ and the state is left unchanged.  Expanding to first order in $\epsilon$, the Kraus operator can be written as
\begin{equation}\label{kraus}
M_{\alpha}^{\kk} = \sqrt{p_{\alpha}^{\kk}} \left[ \mathbb{I} + \epsilon L_{\alpha}^{\kk} \right] + O(\epsilon^2),
\end{equation}
where $L_\alpha$ is a dynamical generator of the system, characterizing the first-order effect of outcome $\alpha$ on the system. The completeness condition requires $\sum_\alpha p_\alpha=1$ and $\sum_\alpha p_\alpha(L_\alpha+L_\alpha^\dagger)=0$.



Although each step barely disturbs the system, the effect of perturbation accumulates along a trajectory.
 For a record of length $m$, the Kraus operator associated with the complete record is
\begin{equation}\label{eq:kraustraj}
    K_{\boldsymbol{\alpha}} = M_{\alpha_m}M_{\alpha_{m-1}} \cdots M_{\alpha_1}\approx \sqrt{p_0(\boldsymbol\alpha)} \left[ I + \epsilon \mathcal{L}_{\boldsymbol\alpha} \right], 
\end{equation}
where 
$ p_0(\boldsymbol{\alpha}) = \prod_{k=1}^m p_{\alpha_k}^{\kk}$ is the record distribution with $\epsilon=0$, and the accumulated generator  $ \mathcal{L}_{\boldsymbol{\alpha}} = \sum_{k=1}^m L_{\alpha_k}^{\kk} $ depicts the effect along the complete record, to the first order. 
 Note that we do not require $\mathcal{L}_{\a}$ to be commute at the first order, since operator-ordering effects enter at $O(\epsilon^2)$~\citep{SM}, while as $m$ increases, the omitted higher-order terms may become non-negligible, limiting the validity of the first-order approximation. 
 
Inserting Eq.~\eqref{eq:kraustraj} into $P_{\text{post}}(\boldsymbol{\alpha})$, the joint probability of the record and successful postselection becomes
\begin{equation}
P_{\text{post}}(\boldsymbol{\alpha}) \approx p_0(\boldsymbol{\alpha}) P_f^{(0)} (1 + 2\epsilon \operatorname{Re}[(\mathcal{L}_{\boldsymbol{\alpha}})_w]),
\end{equation}
where the zeroth-order postselection probability is
$
P_f^{(0)} = \text{Tr}[\Pi_f \rho_i]
$, and 
$
(\mathcal{L}_{\boldsymbol{\alpha}})_w = \text{Tr}[\Pi_f \mathcal{L}_{\boldsymbol{\alpha}} \rho_i]/ \text{Tr}[\Pi_f \rho_i]
$. 
We define $(\mathcal{L}_{\boldsymbol{\alpha}})_w$ as the WV of the measurement record $\boldsymbol{\alpha}$.
This equation gives the general WV response based on the measurement record: postselection modifies the probability of the entire measurement record through the WV of its accumulated dynamical generator~\citep{PhysRevA.96.062110,Ban2015,PhysRevLett.111.160401}. The normalized conditional distribution of the complete record is
$
P_{\rm ps}(\a) = P_{\text{post}}(\boldsymbol{\alpha})/\sum_{\boldsymbol{\alpha}} P_{\text{post}}(\boldsymbol\alpha)
$~\citep{PhysRevA.89.012121}. For any record observable $X(\boldsymbol{\alpha})$, the conditional average after  postselection is
$
\langle X \rangle_{\rm ps} = \sum_{\boldsymbol{\alpha}} X(\boldsymbol{\alpha}) P_{\rm ps}(\boldsymbol{\alpha})
$. 
To first order in the weak response, this gives
\begin{equation}\label{final}
     \langle X \rangle_{\rm ps} \approx \langle X \rangle_0 + 2\epsilon \operatorname{Re} \left[ \langle X(\boldsymbol{\alpha}) (\mathcal{L}_{\boldsymbol{\alpha}})_w \rangle_0 - \langle X \rangle_0 \langle (\mathcal{L}_{\boldsymbol{\alpha}})_w \rangle_0 \right].
\end{equation}
This relationship directly connects the WVs to the statistical averages of repeated measurement records. Eq.~(\ref{final}) 
generalizes the WV response from individual measurement outcomes to arbitrary observables of the complete measurement record. We now specialize this general result to identical commuting binary WMs, for which the complete measurement record can be reduced to a single collective variable.

\paragraph{Physical model.}
The weak measurements of a generic quantum system can be realized by performing RIMs on an ancilla qubit coupled to the system.
In each RIM, the ancilla qubit is firstly initialized in $|0\rangle_{q}$
and subjected to a $\pi/2$ rotation $R_{\varphi_{1}}$, where $R_{\varphi_{1}}(\pi/2)=\exp[-i(\cos\varphi_{1}\sigma^{q}_{x}+\sin\varphi_{1}\sigma^{q}_{y})\pi/4]$. 
This rotation prepares the superposition state $|\psi\rangle_{q}=R_{\varphi_{1}}(\pi/2)|0\rangle_{q}$~\citep{PhysRevA.111.022406}.
Then the system and ancilla interact for a duration $t$ under 
\begin{equation}\label{ham}
H_{\mathrm{int}}= g\,\sigma_z^{q}\otimes \mathcal{A},
\end{equation}
where $g$ is the coupling strength, $\sigma_z^{q}=|0\rangle_q\langle0|-|1\rangle_q\langle1|$, $\mathcal{A}=\mathcal{A}^\dagger$ is the target observable, and $\theta=2gt$ denotes the dimensionless measurement strength.
After an interaction time $t$, the ancilla is subjected to a second Ramsey $\pi/2$ rotation $R_{\varphi_2}(\pi/2)$ and is projectively measured in the basis $\{|0\rangle_q,|1\rangle_q\}$. 
Assigning $ \alpha = \pm 1 $ to the outcomes  $|0\rangle_q $ and  $ |1\rangle_q $, respectively, the corresponding Kraus operator is $M_{\alpha}=\langle \alpha|R_{\varphi_2}(\pi/2)UR_{\varphi_1}(\pi/2)|0\rangle_q =(U_0-\alpha e^{i\Delta\varphi}U_1)/2$, where $U=\exp[-i\theta\sigma_z^q\otimes \mathcal{A}/2]$, $U_0=e^{-i\theta\mathcal{A}/2}$, $U_1=e^{i\theta\mathcal{A}/2}$, and $\Delta\varphi=\varphi_1-\varphi_2$. Up to global phases, the two Kraus operators can be written as $M_{\alpha}=[\cos(\theta\mathcal{A} /2)\, I+\alpha\sin(\theta\mathcal{A}/2)\, ]/\sqrt{2}$~\citep{mzw5-t326,SM}, where we choose $\Delta \varphi=\pi/2$. For $\theta\ll1$, 
we have
$
M_{\alpha}= (I + \alpha\theta \mathcal{A}/2)/\sqrt{2} + O(\theta^2)
$, which is the small-strength expansion of the standard two-outcome WM operator~\citep{Luc2020Quantumness,Ochoa2018Simultaneous}. Thus, $\theta$ controls the response of a symmetric binary WM of $\mathcal{A}$.

Having established the single-step map, we now consider the complete measurement process. For the binary record introduced above, $\alpha_k = \pm 1$ and the zeroth-order probability $p_0(\alpha) = 2^{-m}$~\citep{PhysRevLett.95.110409}. We take $\rho_i = |\psi_i\rangle\langle\psi_i|$ and $\Pi_f = |\psi_f\rangle\langle\psi_f|$. Since all single-step Kraus operators $M_{\alpha_k}$ are functions of the same observable $\mathcal{A}$, they commute, and the full-record Kraus operator $K_{\boldsymbol{\alpha}}$ depends on the record only through  
$S = \sum_{k=1}^m \alpha_k$. For binary records, define the collective polarization $X=S/(2m)=q/m-1/2$, where $q$ counts the $+1$ outcomes and $S=2q-m$.
To first order,
$
K_{\boldsymbol{\alpha}}
\approx
2^{-m/2}
\left[
I+(\theta S \mathcal{A})/2
\right]
+O(\theta^2)
$. The probability is therefore
\begin{equation}\label{prob}
P_{\text{post}}(\boldsymbol{\alpha})\approx\frac{\mathrm{Tr}[\Pi_f\rho_i]}{2^m}
\left(
1+\theta\mathrm{Re}[\mathcal{A}_w]
S
\right),
\end{equation}
where
$
\mathcal{A}_w = \text{Tr}(\Pi_f \mathcal{A} \rho_i)/\text{Tr}(\Pi_f \rho_i) 
$
is the WV. This expression reduces to the standard Aharonov–Albert–Vaidman weak value 
$
\mathcal{A}_w = \langle\psi_f|\mathcal{A}|\psi_i\rangle/\langle\psi_f|\psi_i\rangle
$ for pure pre- and postselection states. 
For this binary instrument, Eq.~(\ref{final}) applies with $\epsilon=\theta$ and $(\mathcal{L}_{\boldsymbol{\alpha}})_w=S\mathcal{A}_w/2$. Since $\langle S\rangle_0=0$ and $\langle S^2\rangle_0=m$, it yields $\langle X\rangle_{\mathrm{ps}}\simeq(\theta/2)\operatorname{Re}[\mathcal{A}_w]$. The displacement relative to the unconditioned mean is therefore $\Delta\langle X\rangle\simeq(\theta/2)(\operatorname{Re}[\mathcal{A}_w]-\langle\mathcal{A}\rangle_i)$, where $\langle\mathcal{A}\rangle_i=\operatorname{Tr}(\mathcal{A}\rho_i)$. Thus, $X$ serves as a collective pointer for the weak-value response within the joint weak-response regime~\citep{PhysRevA.76.044103}.
When $\mathrm{Re}[\mathcal{A}_{w}]$ lies outside the eigenvalue spectrum of the measured observable, the conditional shift can exceed the eigenvalue-calibrated linear-response range, which is a hallmark of WVA~\citep{PhysRevA.107.063706,PhysRevA.92.012119}. 
In the present discrete-time setting, the WVA is encoded not in a single pointer displacement but in the collective statistics of the full measurement record. In the weak-response regime, the mean cumulative signal grows linearly with the number of measurements, whereas the corresponding normalized polarization exhibits an $m$-independent displacement and fluctuations that narrow as $m^{-1/2}$. Repeated WMs therefore enhance the conditional statistical distinguishability of the postselection-induced shift within the successfully postselected ensemble, provided that the accumulated measurement backaction remains perturbative. This collective record perspective identifies polarization as a natural observable through which the WV manifests in sequential measurement dynamics.

\begin{figure}[t]
\centering
\includegraphics[width=\linewidth]{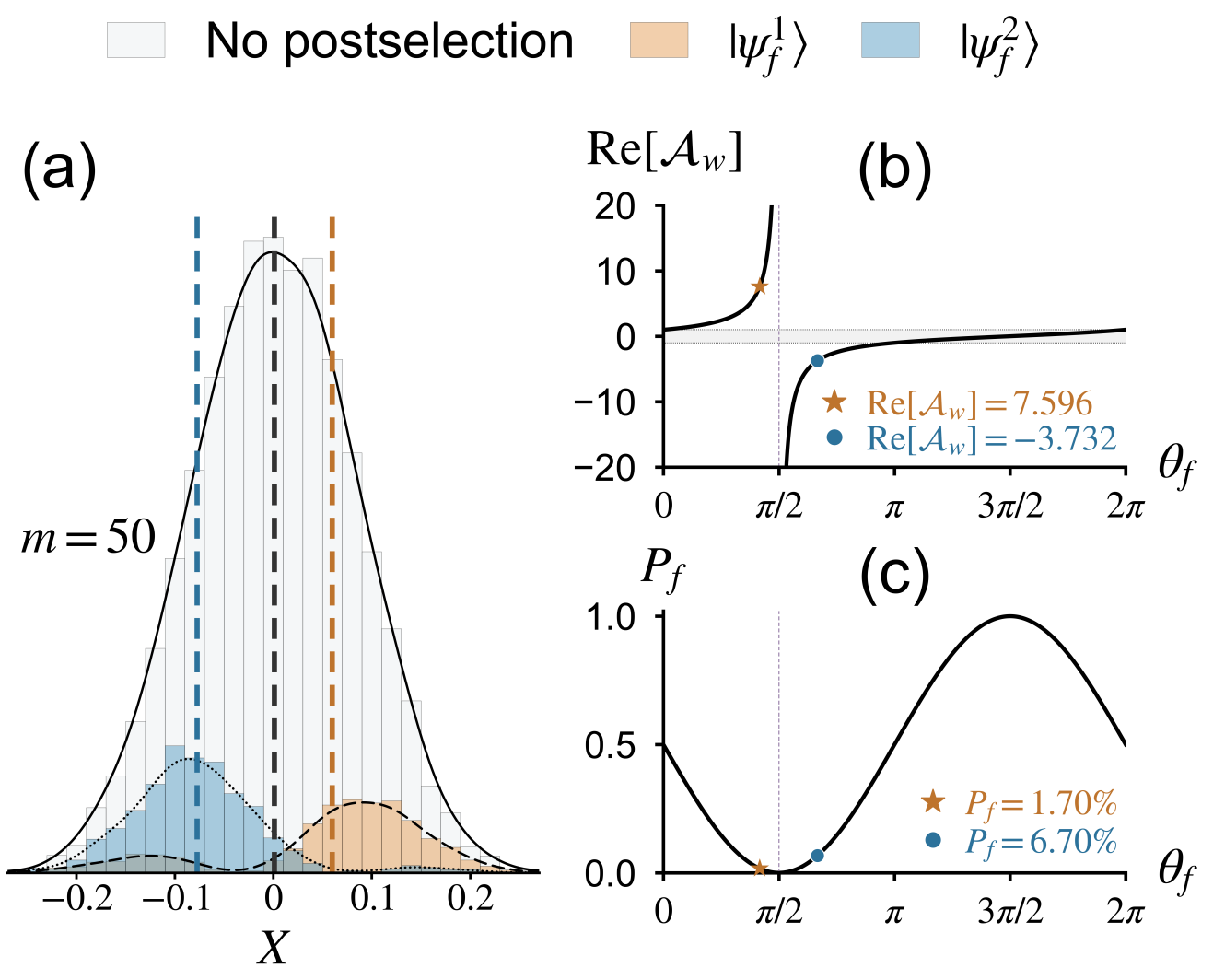} 
\caption{
Qubit record statistics.
(a)  Distributions of the record polarization $X$ for $m=50$ and $\theta=\pi/40$, without postselection (gray) and with postselection onto $|\psi_f^1\rangle$ (orange) and $|\psi_f^2\rangle$ (blue). Dashed lines represent the mean values of the distributions.
(b) Real WV and (c) initial overlap as functions of $\theta_f$ at $\phi_f=\pi$.
The star and circle mark the two postselections, with $\theta_f^1=5\pi/12$ and $\theta_f^2=2\pi/3$. Horizontal dotted lines in (b) indicate the eigenvalue bounds.
We take $10^4$ trajectories per simulation.
\label{fig:2} }
\end{figure}

\paragraph{Example I.}

We now specialize to $\mathcal A=\sigma_z$, for which $\mathcal A^2=I$.
A general qubit input and rank-one postselection are described by
$\rho_i=(I+\mathbf n\cdot\boldsymbol\sigma)/2$ and
$\Pi_f=(I+\mathbf r\cdot\boldsymbol\sigma)/2$, respectively.
Here, $\mathbf n$ and $\mathbf r$ are the corresponding Bloch vectors,
with $|\mathbf n|\leq1$ and $|\mathbf r|=1$.
For $P_f^{(0)}>0$, the WV becomes
\begin{equation}
\mathcal A_w=\frac{\text{Tr}[\Pi_f \sigma_z \rho_i]}{\text{Tr}[\Pi_f \rho_i] }
=
\frac{n_z+r_z+i(\mathbf n\times\mathbf r)_z}
     {1+\mathbf n\cdot\mathbf r}.
\end{equation}

Within the weak-response regime, $\operatorname{Re}[\mathcal A_w]$
governs the leading bias of the collective measurement record.
For the finite-strength example in Fig.~\ref{fig:2}, we specialize this general setting to the
balanced input $|\psi_i\rangle=(|0\rangle+|1\rangle)/\sqrt2$,
corresponding to $\mathbf n=(1,0,0)$.
The postselected states are parameterized as
$|\psi_f^j\rangle=\cos(\theta_f^j/2)|0\rangle
+e^{i\phi_f^j}\sin(\theta_f^j/2)|1\rangle$,
with $\mathbf r_j=(\sin\theta_f^j\cos\phi_f^j,
\sin\theta_f^j\sin\phi_f^j,\cos\theta_f^j)$.
In Fig.~\ref{fig:2}, we set $\phi_f^1=\phi_f^2=\pi$,
$\theta_f^1=5\pi/12$, and $\theta_f^2=2\pi/3$.
These choices give
$\operatorname{Re}[\mathcal A_w^1]\simeq7.596$,
$\operatorname{Re}[\mathcal A_w^2]\simeq-3.732$, and
$\operatorname{Im}[\mathcal A_w^1]
=\operatorname{Im}[\mathcal A_w^2]=0$.
For identical commuting measurements, all records with $q$ positive outcomes share the same operator $K(q)$ and have multiplicity $\binom mq$. The exact conditional probability is
\begin{equation}
P_{\rm ps}(q)=\binom mq
\frac{|\langle\psi_f|K(q)|\psi_i\rangle|^2}{P_f(m,\theta)}.
\label{eq:exact-conditional-record}
\end{equation}
Here $P_f(m,\theta)=\operatorname{Tr}[\Pi_f\Phi^m(\rho_i)]$ is the probability of successful final-state postselection after the complete sequence. For the balanced qubit, define $D_m=1+r_x(\cos\theta)^m$, so that $P_f(m,\theta)=D_m/2$. The exact conditional mean is $\langle X\rangle_{\rm ps}={r_z\sin\theta}/{2D_m}$
and its variance is $\operatorname{Var}_{\rm ps}(X) ={1}/(4m)+{(m-1)\sin^2\theta}/(4mD_m) -[r_z\sin\theta/(2D_m)]^2.$

Because $S=2q-m$, normalizing Eq. (\ref{prob}) over successfully postselected records and summing over the $\binom mq$ records with $q$ positive outcomes gives
$
P_{\rm ps}(q) \approx 2^{-m} {m\choose q} \left[ 1+ \theta\operatorname{Re}[\mathcal{A}_w](2q-m) \right]
$. Consequently, 
$
\langle X\rangle_{\text{ps}} \approx \theta\operatorname{Re}[\mathcal{A}_w]/2
$. 
Therefore, the leading conditional displacement is 
$
\Delta\langle X\rangle
=\langle X\rangle_{\text{ps}}
-\langle X\rangle_{\text{no}}
\approx \theta \operatorname{Re} [\mathcal{A}_w]/2$. This expression shows that the real part of the WV acts as an amplification factor for the conditional shift in the polarization distribution. The record expansion is controlled by the accumulated response. For typical balanced records, $S=O(\sqrt m)$, so the linear conditional correction is small when $\sqrt m\,|\theta\mathcal{A}_w|\ll1$. We also require $m\theta^2\ll1$ to keep the accumulated qubit measurement backaction perturbative~\citep{Coto2017}. 

The dominant polarization values are those observed most frequently in the postselected ensemble. They maximize $P_{\rm ps}(q)$, or equivalently minimize $-\ln P_{\rm ps}(q)$.
Within the joint weak-response window, a binomial approximation for $q$ reproduces the leading bias induced by postselection.
Its effective probability for a positive outcome is $p_{\rm ps}=(1+\theta\operatorname{Re}[\mathcal{A}_w])/2$.
For this binomial distribution, the peak polarization and mean differ by at most one grid spacing, $1/m$, which decreases with sequence length.
Beyond this approximation, the peak and mean must be evaluated separately from the exact conditional distribution.

Figure~\ref{fig:2}(a) compares unconditioned records (gray) with records retained by postselection onto $|\psi_f^1\rangle$ (orange) or $|\psi_f^2\rangle$ (blue). 
For the settings in Fig.~\ref{fig:2}(a), $\langle X\rangle_{\rm ps}$ gives shifts $0.05895$ and $-0.07607$, respectively, compared with the linear predictions $0.29828$ and $-0.14656$. 
 A controlled comparison at weaker coupling, with the initial and final states fixed and the sequence length varied, gives relative deviations below $1.63\%$ for the mean displacement and below $1.59\%$ for the leading width across all four postselection settings; details are given in the Supplemental Material~\citep{SM}. Only in the unimodal perturbative regime can the mean displacement and saddle-point position both be approximated by $\theta\operatorname{Re}[\mathcal{A}_w]/2$.
This relation is the record-space counterpart of the conventional weak-pointer shift $\langle x\rangle_{\text{ps}}\approx \epsilon\operatorname{Re}[\mathcal{A}_w]$~\cite{Dressel2014}, with $X$ as an effective collective-pointer variable.
Figures~\ref{fig:2}(b,c) show how the WV and initial overlap vary with the postselection angle. The marked postselections yield WVs $7.596$ and $-3.732$, outside the spectrum $[-1,1]$. As the pre- and postselected states approach orthogonality, a larger formal WV accompanies a smaller initial overlap. Panel (c), labeled $P_f$, shows this overlap $P_f^{(0)}=|\langle\psi_f|\psi_i\rangle|^2$, with marked values $1.70\%$ and $6.70\%$. The exact sequence success probabilities $P_f(m,\theta)$ are instead $8.61\%$ and $12.89\%$, because accumulated backaction changes the final state.
Within the joint weak-response window, the leading displacement is independent of $m$, whereas the conditional width is $\sigma_X\approx1/(2\sqrt m)$. We quantify their ratio by
\begin{equation}
\mathcal R_{\rm ps}^{\rm cond} =\frac{|\langle X\rangle_{\rm ps}-\langle X\rangle_{\rm no}|}{\sigma_X} \approx\theta\big|\operatorname{Re}[\mathcal A_w]-\langle\mathcal A\rangle_i\big|\sqrt m.
\label{eq:conditional-displacement-width}
\end{equation}
This ratio compares the shift with the spread of a single successful record. For $N_{\rm succ}$ independent successful records, the standard error of their sample mean is $\sigma_X/\sqrt{N_{\rm succ}}$; the uncertainty of a measured shift also depends on the success probability and on how the unconditioned reference is estimated.

\paragraph{Example II.}
We next consider a single bosonic mode with photon-number observable $\mathcal{A}=\hat a^\dagger\hat a$. The $n$-photon Fock state $|n\rangle$ satisfies $\mathcal A|n\rangle=n|n\rangle$.
For arbitrary normalized pure pre- and postselected states
$|\psi_i\rangle=\sum_{n=0}^{\infty}c_n|n\rangle$ and
$|\psi_f\rangle=\sum_{n=0}^{\infty}d_n|n\rangle$, the coefficients $c_n=\langle n|\psi_i\rangle$ and $d_n=\langle n|\psi_f\rangle$ are their complex probability amplitudes in the Fock basis, satisfying $\sum_n|c_n|^2=\sum_n|d_n|^2=1$. Their squared moduli, $|c_n|^2$ and $|d_n|^2$, give the corresponding photon-number probabilities. Provided $\langle\psi_f|\psi_i\rangle\ne0$, the corresponding WV is
\begin{equation}
\mathcal{A}_w
=
\frac{\langle\psi_f|\mathcal{A}|\psi_i\rangle}
{\langle\psi_f|\psi_i\rangle}
=
 \frac{\sum_n^{\infty} n d_n^{*}c_n}
{\sum_n^{\infty} d_n^{*}c_n}.
\end{equation}

We choose the vacuum--Fock superposition $|\psi_i\rangle=|\Psi(4)\rangle=(|0\rangle+|4\rangle)/\sqrt{2}$, so that $c_0=c_4=1/\sqrt{2}$, all other $c_n$ vanish, and $\langle\mathcal A\rangle_i=2$. States in the family $|\Psi(N)\rangle=(|0\rangle+|N\rangle)/\sqrt{2}$ have been prepared for single-mode quantum metrology in a superconducting circuit~\cite{Wang2019SingleMode}. We consider the two postselection states $|\psi_f^1\rangle=\cos(\theta_f/2)|0\rangle+e^{i\phi_f}\sin(\theta_f/2)[\cos(\vartheta_f/2)|1\rangle+\sin(\vartheta_f/2)|4\rangle]$ and $|\psi_f^2\rangle=\mathcal N_\chi(|\alpha\rangle+e^{i\chi}|-\alpha\rangle)$, where $\theta_f$ and $\vartheta_f$ control the relative amplitudes in the first state, $\phi_f$ is its relative phase, $\mathcal N_\chi=[2(1+e^{-2|\alpha|^2}\cos\chi)]^{-1/2}$ normalizes the cat state, $\alpha$ is the complex coherent-state amplitude, and $\chi$ is the relative phase between its two coherent components. The number-diagonal Kraus operators preserve the input support $\operatorname{span}\{|0\rangle,|4\rangle\}$. Since only $c_0$ and $c_4$ are nonzero, the WV for either postselection reduces to $\mathcal A_w=4d_4^*/(d_0^*+d_4^*)$.
Here $d_0=\langle0|\psi_f\rangle$ and $d_4=\langle4|\psi_f\rangle$ are the vacuum and four-photon amplitudes of the chosen postselected state. Although only these two amplitudes enter the reduced WV, the full postselected state retains its normalization. The overlap probability before the measurement sequence is $P_f^{(0)}=|\langle\psi_f|\psi_i\rangle|^2=|d_0^*+d_4^*|^2/2$. In general, $\mathcal A_w^j=\operatorname{Re}[\mathcal A_w^j]+i\operatorname{Im}[\mathcal A_w^j]$, where $\operatorname{Re}$ and $\operatorname{Im}$ denote the real and imaginary parts, respectively, and $j=1,2$ labels the two postselections. Within the controlled weak-response window, the leading postselection-induced polarization shift is $\Delta\langle X\rangle\approx\theta(\operatorname{Re}[\mathcal A_w]-2)/2$, where $\theta=2gt$ and $\Delta\langle X\rangle=\langle X\rangle_{\rm ps}-\langle X\rangle_{\rm no}$.

For the first postselection, the relevant amplitudes are $d_0=\cos(\theta_f/2)$ and $d_4=e^{i\phi_f}\sin(\theta_f/2)\sin(\vartheta_f/2)$. Setting $\theta_f=\vartheta_f=\pi/2$ and $\phi_f=\pi$ gives $d_0=1/\sqrt{2}$ and $d_4=-1/2$. For the even-cat postselection with $\chi=0$, the corresponding amplitudes are $d_0=2\mathcal N_0e^{-|\alpha|^2/2}$ and $d_4=d_0\alpha^4/\sqrt{24}$, where $\mathcal N_0$ is $\mathcal N_\chi$ at $\chi=0$. Hence $\mathcal A_w^2=4(\alpha^*)^4/[\sqrt{24}+(\alpha^*)^4]$. Choosing $\alpha=e^{i\pi/4}(5\sqrt{24}/3)^{1/4}$ gives $\operatorname{Re}[\mathcal A_w^1]=-4(1+\sqrt2)\simeq-9.657$, $\operatorname{Re}[\mathcal A_w^2]=10$, and $\operatorname{Im}[\mathcal A_w^1]=\operatorname{Im}[\mathcal A_w^2]=0$.
Both WVs are therefore purely real and lie outside the interval $[0,4]$ spanned by the occupied number sectors; the first also lies below the lower bound of the full photon-number spectrum.

\begin{figure}[t]
\centering
\includegraphics[width=\linewidth]{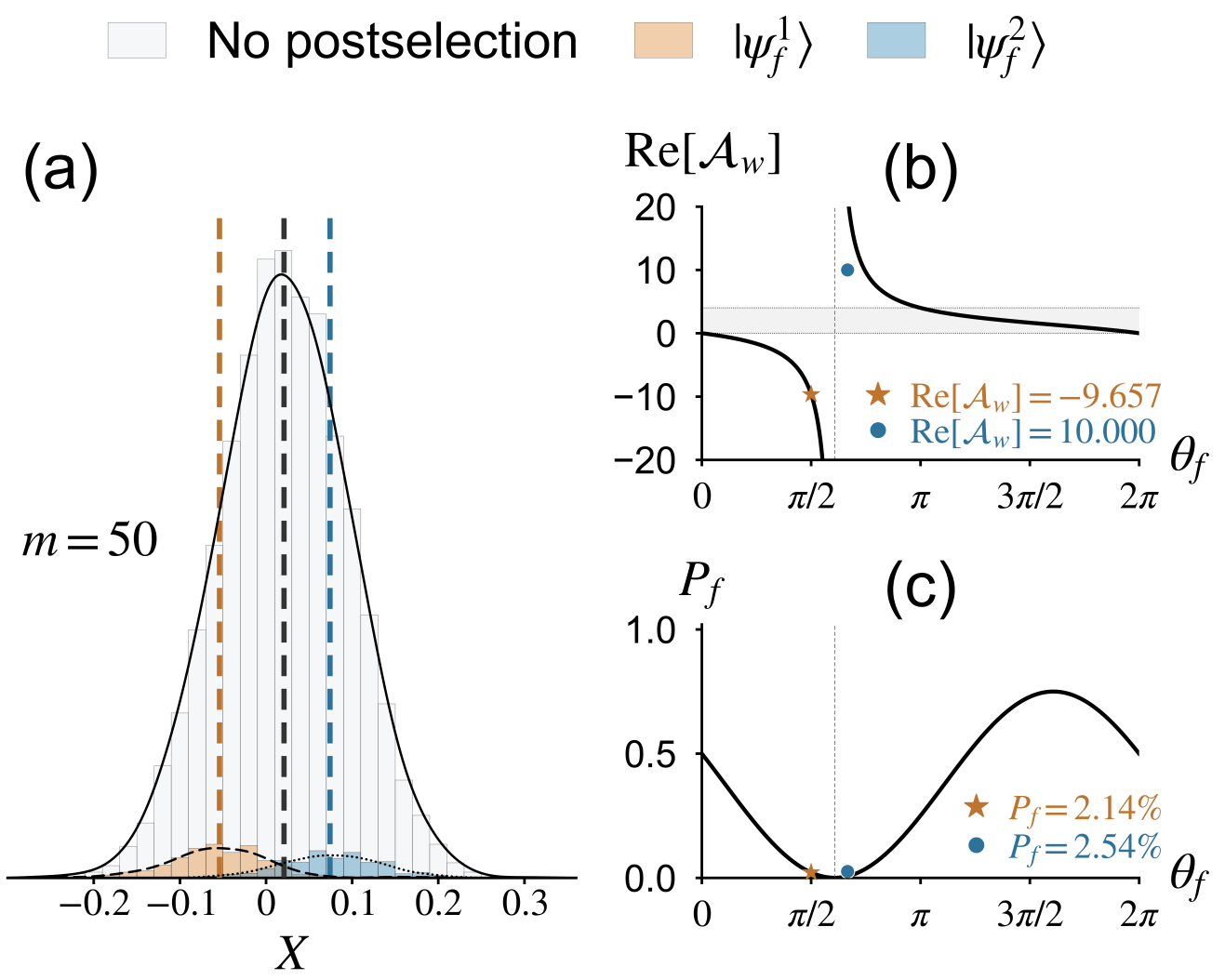} 
\caption{
Oscillator record statistics for the preselected state
$|\psi_i\rangle$
(a) Unnormalized counts without postselection (gray) and with postselection of states $|\psi_f^1\rangle$ (orange) and $|\psi_f^2\rangle$ (blue). Dashed lines represent the mean values of the distributions.
(b) Real WV and (c) initial overlap as functions of  $\theta_f$.
Blue circles show the fixed cat-state values
$\operatorname{Re}[\mathcal A_w^2]=10$ and
$P_{f,2}^{(0)}=2.54\%$.
Orange stars mark the first postselection in (a).
The blue circles are placed at $\theta_f=2\pi/3$ for comparison;
this postselection is independent of the angular scan.
Horizontal dotted lines in (b)
mark the bounds $0$ and $4$ of the occupied number sectors.
The purple dashed vertical line marks the initial orthogonality angle
$\theta_f=2\arctan(\sqrt{2})\simeq1.911$.
Parameters: $m=50$, $t=\pi/300$, $g=1$, $\vartheta_f=\pi/2$,
$\phi_f=\pi$, $\alpha=e^{i\pi/4}(5\sqrt{24}/3)^{1/4}$, and $\chi=0$; in (a),
$\theta_f=\pi/2$, with $10^4$ initial trajectories shared by the
two postselections.
Initial overlap and sequence success are distinct quantities.
}
\label{fig:3}
\end{figure}

Figures~\ref{fig:3}(b,c) compare the real parts of the WVs and the initial overlaps of the two postselections. The black curves describe the first postselection as $\theta_f$ varies, with orange stars marking $\theta_f=\pi/2$. The cat state is independent of $\theta_f$. Blue circles, placed at $\theta_f=2\pi/3$ for comparison, show its WV and initial overlap. At the selected settings, the initial overlap probabilities are $P_{f,1}^{(0)}\simeq2.145\%$ and $P_{f,2}^{(0)}\simeq2.543\%$. After the complete sequence, the exact success probabilities are $P_{f,1}(m,\theta)\simeq3.662\%$ and $P_{f,2}(m,\theta)\simeq3.362\%$.

Figure~\ref{fig:3}(a) shows the finite-strength record distributions for $m=50$ and $gt=\pi/300$. The record polarization is $X=q/m-1/2$, where $q$ counts the $+$ outcomes in the $m$ measurements. The exact unconditioned mean is $\langle X\rangle_{\rm no}=0.02092$, and the conditional means are $\langle X\rangle_{{\rm ps},1}=-0.05082$ and $\langle X\rangle_{{\rm ps},2}=0.08402$. The two postselections thus produce displacements of $-0.07174$ and $+0.06310$ relative to the unconditioned mean. The leading weak-response formula instead predicts $-0.12207$ and $+0.08378$, respectively. These differences place the captioned settings outside the controlled weak-response window, and the distributions in panel (a) are evaluated using the exact finite-sequence calculation.

\paragraph{Experimental considerations.}
Finally, we discuss the feasibility of our scheme in present-day experimental platforms. The protocol requires the following two ingredients: (i) sequential RIMs, including repeated ancilla-target couplings of the form of Eq.~(\ref{ham}), each followed by ancilla readout; and (ii) a final projective measurement of the target for postselection. The first ingredient is available in a broad class of platforms. For example, sequential weak measurements of a target $^{14}$N nuclear spin via the RIM on electron spin have been demonstrated through the hyperfine coupling \cite{Blok2014,Liu2017}, which has been used for tracking the precession of a single nuclear spin \cite{Pfender2019NatCommun,Cujia2019Tracking}. In circuit QED, the ancilla-target coupling is realized by the dispersive coupling between a transmon ancilla and a cavity mode, as exploited in repeated photon-parity tracking~\cite{sun2014tracking,Ofek2016Nature}, while high-fidelity transmon rotations and single-shot dispersive readout provide the remaining elements of each RIM~\cite{Blais2021}.

Ingredient (ii) can be realized by a unitary transformation followed by a projective measurement of the target in a fixed basis. For example, one can apply $U_f^\dagger$ with $U_f\ket{0}=\ket{\psi_f}$, then selecting the trajectories with measurement outcome $\ket{0}$ yields the postselection with final state $\ket{\psi_f}$. Transmon and trapped-ion targets can be read out directly, whereas nuclear spins and cavity modes are read out through the ancilla. For the oscillator, $U_f^\dagger$ can be implemented with selective number-dependent arbitrary phase and displacement gates~\cite{Heeres2015PRL,Krastanov2015} or echoed conditional displacements~\cite{Eickbusch2022}, followed by vacuum-selective ancilla pulse and readout~\cite{Johnson2010QND}.  The two ingredients have already been combined in single-step ($m=1$) experiments: an ancilla-based weak measurement followed by a final projective readout of the target has revealed nonclassical WVs in superconducting circuits with the target being another transmon~\cite{Groen2013} and tenfold WVA for a nuclear spin in diamond~\cite{Blok2014}. Weak-to-strong transition of quantum measurement together with a postselection has also been demonstrated in a trapped-ion system \cite{pan2020weak}.

\paragraph{Conclusions and outlooks.} We have shown that final-state postselection reveals WVs in sequential WMs by reweighting complete measurement records. 
The leading reweighting is governed by the WV of the accumulated dynamical generator. For identical symmetric binary WMs of a fixed observable, commutativity reduces the record dependence to the collective record polarization $X$. This variable therefore serves as an effective pointer, with leading displacement $\Delta\langle X\rangle\simeq(\theta/2)(\operatorname{Re}[\mathcal A_w]-\langle\mathcal A\rangle_i)$ relative to the unconditioned mean. Within the weak-measurement regime, this $m$-independent displacement and the $m^{-1/2}$ width yield a conditional displacement-to-width ratio proportional to $\sqrt m$. Exact qubit and oscillator distributions delimit this regime as measurement backaction accumulates.

The present framework can be extended in several directions. First, the finite-sequence analysis could be extended beyond first order to noncommuting Kraus operators. This would clarify how temporal correlations in ordered records modify the WV response beyond that captured by the collective polarization~\citep{394k-lj6s,Cech2025,Cech2026}. Second, experiments in circuit quantum electrodynamics and trapped ions could test the predicted polarization shift and quantify its sensitivity to readout errors and accumulated measurement backaction. Finally, its metrological potential should be evaluated against an unconditioned protocol at fixed total measurement resources, accounting for failed postselection events.

\paragraph{Acknowledgments.}
This work was supported by the National Natural Science Foundation of China (Grants No. 12574082, No. 12174379, and No. E31Q02BG), the Chinese Academy of Sciences (Grants No. E0SEBB11 and No. E27RBB11), Quantum Science and Technology National Science and Technology Major Project (Grant No. 2021ZD0302300), and the Chinese Academy of Sciences Project for Young Scientists in Basic Research
(Grant No. YSBR-090).

\paragraph{Data Availability.}
Analytical distributions and parameters are provided in the main text and Supplemental Material. The code used to generate the figures will be made available in Ref.~\citep{janaGitHub2026}. 

\bibliography{refs}

@article{PhysRevA.107.012217,
  title = {Sequential generalized measurements: Asymptotics, typicality, and emergent projective measurements},
  author = {Ma, Wen-Long and Li, Shu-Shen and Liu, Ren-Bao},
  journal = {Phys. Rev. A},
  volume = {107},
  issue = {1},
  pages = {012217},
  numpages = {10},
  year = {2023},
  month = {Jan},
  publisher = {American Physical Society},
  doi = {10.1103/PhysRevA.107.012217},
  url = {https://link.aps.org/doi/10.1103/PhysRevA.107.012217}
}

@article{PhysRevA.81.012103,
  title = {Complete characterization of post-selected quantum statistics using weak measurement tomography},
  author = {Hofmann, Holger F.},
  journal = {Phys. Rev. A},
  volume = {81},
  issue = {1},
  pages = {012103},
  numpages = {5},
  year = {2010},
  month = {Jan},
  publisher = {American Physical Society},
  doi = {10.1103/PhysRevA.81.012103},
  url = {https://link.aps.org/doi/10.1103/PhysRevA.81.012103}
}

@article{66yg-qs8x,
  title = {Spectrum measurement of quantum channels and application to Hamiltonian parameter estimation},
  author = {Jin, Yuan-De and Ma, Wen-Long},
  journal = {Phys. Rev. A},
  volume = {113},
  issue = {1},
  pages = {L010406},
  numpages = {7},
  year = {2026},
  month = {Jan},
  publisher = {American Physical Society},
  doi = {10.1103/66yg-qs8x},
  url = {https://link.aps.org/doi/10.1103/66yg-qs8x}
}

@article{PhysRevLett.134.080802,
  title = {Enhancing Interferometry Using Weak Value Amplification with Real Weak Values},
  author = {Huang, Jing-Hui and Jordan, Kyle M. and Dada, Adetunmise C. and Hu, Xiang-Yun and Lundeen, Jeff. S.},
  journal = {Phys. Rev. Lett.},
  volume = {134},
  issue = {8},
  pages = {080802},
  numpages = {6},
  year = {2025},
  month = {Feb},
  publisher = {American Physical Society},
  doi = {10.1103/PhysRevLett.134.080802},
  url = {https://link.aps.org/doi/10.1103/PhysRevLett.134.080802}
}

@article{PhysRevA.109.022224,
  title = {Origin of meter fluctuations in weak measurement interactions},
  author = {Matsushita, Tomonori and Hofmann, Holger F.},
  journal = {Phys. Rev. A},
  volume = {109},
  issue = {2},
  pages = {022224},
  numpages = {12},
  year = {2024},
  month = {Feb},
  publisher = {American Physical Society},
  doi = {10.1103/PhysRevA.109.022224},
  url = {https://link.aps.org/doi/10.1103/PhysRevA.109.022224}
}

@article{PhysRevLett.60.1351,
  title = {How the result of a measurement of a component of the spin of a spin-1/2 particle can turn out to be 100},
  author = {Aharonov, Yakir and Albert, David Z. and Vaidman, Lev},
  journal = {Phys. Rev. Lett.},
  volume = {60},
  issue = {14},
  pages = {1351--1354},
  numpages = {0},
  year = {1988},
  month = {Apr},
  publisher = {American Physical Society},
  doi = {10.1103/PhysRevLett.60.1351},
  url = {https://link.aps.org/doi/10.1103/PhysRevLett.60.1351}
}

@article{394k-lj6s,
  title = {Dynamical discontinuities in repeated weak measurements revealed by complex weak values},
  author = {Ferraz, Lorena Ballesteros},
  journal = {Phys. Rev. A},
  volume = {113},
  issue = {4},
  pages = {042413},
  numpages = {10},
  year = {2026},
  month = {Apr},
  publisher = {American Physical Society},
  doi = {10.1103/394k-lj6s},
  url = {https://link.aps.org/doi/10.1103/394k-lj6s}
}

@article{PhysRevLett.117.170402,
  title = {Measuring Incompatible Observables by Exploiting Sequential Weak Values},
  author = {Piacentini, F. and Avella, A. and Levi, M. P. and Gramegna, M. and Brida, G. and Degiovanni, I. P. and Cohen, E. and Lussana, R. and Villa, F. and Tosi, A. and Zappa, F. and Genovese, M.},
  journal = {Phys. Rev. Lett.},
  volume = {117},
  issue = {17},
  pages = {170402},
  numpages = {6},
  year = {2016},
  month = {Oct},
  publisher = {American Physical Society},
  doi = {10.1103/PhysRevLett.117.170402},
  url = {https://link.aps.org/doi/10.1103/PhysRevLett.117.170402}
}

@article{PhysRevLett.95.110409,
  title = {Weak Measurements Are Universal},
  author = {Oreshkov, Ognyan and Brun, Todd A.},
  journal = {Phys. Rev. Lett.},
  volume = {95},
  issue = {11},
  pages = {110409},
  numpages = {4},
  year = {2005},
  month = {Sep},
  publisher = {American Physical Society},
  doi = {10.1103/PhysRevLett.95.110409},
  url = {https://link.aps.org/doi/10.1103/PhysRevLett.95.110409}
}

@article{Kedem2010,
  title = {Modular Values and Weak Values of Quantum Observables},
  author = {Kedem, Y. and Vaidman, L.},
  journal = {Phys. Rev. Lett.},
  volume = {105},
  issue = {23},
  pages = {230401},
  numpages = {4},
  year = {2010},
  month = {Nov},
  publisher = {American Physical Society},
  doi = {10.1103/PhysRevLett.105.230401},
  url = {https://link.aps.org/doi/10.1103/PhysRevLett.105.230401}
}

@article{Davies1970,
  author  = {Davies, E. B. and Lewis, J. T.},
  title   = {An operational approach to quantum probability},
  journal = {Commun. Math. Phys.},
  volume  = {17},
  pages   = {239},
  year    = {1970},
  doi     = {10.1007/BF01647093}
}

@article{Ozawa1984,
  author  = {Ozawa, Masanao},
  title   = {Quantum measuring processes of continuous observables},
  journal = {J. Math. Phys.},
  volume  = {25},
  pages   = {79},
  year    = {1984},
  doi     = {10.1063/1.526000}
}

@article{Magana-Loaiza2014,
  title = {Amplification of Angular Rotations Using Weak Measurements},
  author = {Maga{\~n}a-Loaiza, Omar S. and Mirhosseini, Mohammad and Rodenburg, Brandon and Boyd, Robert W.},
  journal = {Phys. Rev. Lett.},
  volume = {112},
  issue = {20},
  pages = {200401},
  numpages = {5},
  year = {2014},
  month = {May},
  publisher = {American Physical Society},
  doi = {10.1103/PhysRevLett.112.200401},
  url = {https://link.aps.org/doi/10.1103/PhysRevLett.112.200401}
}

@article{Liu2017,
  title = {Single-Shot Readout of a Nuclear Spin Weakly Coupled to a Nitrogen-Vacancy Center at Room Temperature},
  author = {Liu, Gang-Qin and Xing, Jian and Ma, Wen-Long and Wang, Ping and Li, Chang-Hao and Po, Hoi Chun and Zhang, Yu-Ran and Fan, Heng and Liu, Ren-Bao and Pan, Xin-Yu},
  journal = {Phys. Rev. Lett.},
  volume = {118},
  issue = {15},
  pages = {150504},
  numpages = {5},
  year = {2017},
  month = {Apr},
  publisher = {American Physical Society},
  doi = {10.1103/PhysRevLett.118.150504},
  url = {https://link.aps.org/doi/10.1103/PhysRevLett.118.150504}
}

@article{Blok2014,
  author  = {Blok, M. S. and Bonato, C. and Markham, M. L. and Twitchen, D. J. and Dobrovitski, V. V. and Hanson, R.},
  title   = {Manipulating a qubit through the backaction of sequential partial measurements and real-time feedback},
  journal = {Nat. Phys.},
  volume  = {10},
  pages   = {189},
  year    = {2014},
  doi     = {10.1038/nphys2881}
}

@article{Blais2021,
  author  = {Blais, A. and Grimsmo, A. L. and Girvin, S. M. and Wallraff, A.},
  title   = {Circuit quantum electrodynamics},
  journal = {Rev. Mod. Phys.},
  volume  = {93},
  pages   = {025005},
  year    = {2021},
  doi     = {10.1103/RevModPhys.93.025005}
}

@misc{janaGitHub2026,
  author = {Yuanbek, Janarbek and Jin, Yuan-De and Ma, Wen-Long},
  title  = {Weak Values from Discrete-Time Quantum Trajectories},
  year   = {2026},
  note   = {GitHub repository},
  url    = {https://github.com/yuanjanarbek-source/Weak-Values-from-Discrete-Time-Quantum-Trajectories}
}

@misc{SM,
  note = {See Supplemental Material for detailed derivations of the weak-instrument response, the binary weak-measurement model, the conditional system--ancilla evolution, and the qubit and oscillator examples.}
}

@article{Ban2015,
  author    = {Ban, Masashi},
  title     = {Conditional average in a quantum system with postselection},
  journal   = {Quantum Stud. Math. Found.},
  year      = {2015},
  month     = {sep},
  volume    = {2},
  number    = {3},
  pages     = {263--273},
  doi       = {10.1007/s40509-015-0043-9},
  url       = {https://doi.org/10.1007/s40509-015-0043-9},
  issn      = {2196-5617}
}

@article{PhysRevA.103.062206,
  title = {Experimental test of sequential weak measurements for certified quantum randomness extraction},
  author = {Foletto, Giulio and Padovan, Matteo and Avesani, Marco and Tebyanian, Hamid and Villoresi, Paolo and Vallone, Giuseppe},
  journal = {Phys. Rev. A},
  volume = {103},
  issue = {6},
  pages = {062206},
  numpages = {9},
  year = {2021},
  month = {Jun},
  publisher = {American Physical Society},
  doi = {10.1103/PhysRevA.103.062206},
  url = {https://link.aps.org/doi/10.1103/PhysRevA.103.062206}
}

@article{GeorgievCohen2018,
  title = {Probing finite coarse-grained virtual Feynman histories with sequential weak values},
  author = {Georgiev, Danko and Cohen, Eliahu},
  journal = {Phys. Rev. A},
  volume = {97},
  issue = {5},
  pages = {052102},
  numpages = {11},
  year = {2018},
  month = {May},
  publisher = {American Physical Society},
  doi = {10.1103/PhysRevA.97.052102},
  url = {https://link.aps.org/doi/10.1103/PhysRevA.97.052102}
}

@article{wen2023demonstration,
  title={Demonstration of the quantum principle of least action with single photons},
  author={Wen, Yong-Li and Wang, Yunfei and Tian, Li-Man and Zhang, Shanchao and Li, Jianfeng and Du, Jing-Song and Yan, Hui and Zhu, Shi-Liang},
  journal={Nat. Photonics},
  volume={17},
  number={8},
  pages={717--722},
  year={2023},
  publisher={Nature Publishing Group UK London},
  url = {https://doi.org/10.1038/s41566-023-01212-1},
  doi = {10.1038/s41566-023-01212-1}
}

@article{Aharonov1988,
  title = {How the result of a measurement of a component of the spin of a spin-1/2 particle can turn out to be 100},
  author = {Aharonov, Yakir and Albert, David Z. and Vaidman, Lev},
  journal = {Phys. Rev. Lett.},
  volume = {60},
  issue = {14},
  pages = {1351--1354},
  numpages = {0},
  year = {1988},
  month = {Apr},
  publisher = {American Physical Society},
  doi = {10.1103/PhysRevLett.60.1351},
  url = {https://link.aps.org/doi/10.1103/PhysRevLett.60.1351}
}

@article{Ochoa2018Simultaneous,
  author  = {Ochoa, Maicol A. and Belzig, Wolfgang and Nitzan, Abraham},
  title   = {Simultaneous weak measurement of non-commuting observables: a generalized {Arthurs-Kelly} protocol},
  journal = {Sci. Rep.},
  volume  = {8},
  number  = {1},
  pages   = {15781},
  year    = {2018},
  doi     = {10.1038/s41598-018-33562-0},
  url     = {https://doi.org/10.1038/s41598-018-33562-0}
}

@article{Luc2020Quantumness,
  author  = {Luc, Joanna},
  title   = {Quantumness of States and Unitary Operations},
  journal = {Found. Phys.},
  volume  = {50},
  number  = {11},
  pages   = {1645--1685},
  year    = {2020},
  doi     = {10.1007/s10701-020-00391-z},
  url     = {https://doi.org/10.1007/s10701-020-00391-z}
}

@article{mzw5-t326,
  title = {Verification of contextuality by noncontextual experiments},
  author = {Thio, Jonathan J. and Salmon, Wilfred and Barnes, Crispin H. W. and De Bi\`evre, Stephan and Arvidsson-Shukur, David R. M.},
  journal = {Phys. Rev. A},
  volume = {112},
  issue = {6},
  pages = {062230},
  numpages = {15},
  year = {2025},
  month = {Dec},
  publisher = {American Physical Society},
  doi = {10.1103/mzw5-t326},
  url = {https://link.aps.org/doi/10.1103/mzw5-t326}
}

@article{Dressel2014,
  title = {Colloquium: Understanding quantum weak values: Basics and applications},
  author = {Dressel, Justin and Malik, Mehul and Miatto, Filippo M. and Jordan, Andrew N. and Boyd, Robert W.},
  journal = {Rev. Mod. Phys.},
  volume = {86},
  issue = {1},
  pages = {307--316},
  numpages = {0},
  year = {2014},
  month = {Mar},
  publisher = {American Physical Society},
  doi = {10.1103/RevModPhys.86.307},
  url = {https://link.aps.org/doi/10.1103/RevModPhys.86.307}
}

@article{PhysRevA.94.010103,
  title = {Structural features of sequential weak measurements},
  author = {Di\'osi, Lajos},
  journal = {Phys. Rev. A},
  volume = {94},
  issue = {1},
  pages = {010103(R)},
  numpages = {4},
  year = {2016},
  month = {Jul},
  publisher = {American Physical Society},
  doi = {10.1103/PhysRevA.94.010103},
  url = {https://link.aps.org/doi/10.1103/PhysRevA.94.010103}
}

@article{PhysRevA.76.062105,
  title = {Sequential weak measurement},
  author = {Mitchison, Graeme and Jozsa, Richard and Popescu, Sandu},
  journal = {Phys. Rev. A},
  volume = {76},
  issue = {6},
  pages = {062105},
  numpages = {9},
  year = {2007},
  month = {Dec},
  publisher = {American Physical Society},
  doi = {10.1103/PhysRevA.76.062105},
  url = {https://link.aps.org/doi/10.1103/PhysRevA.76.062105}
}

@book{Wilde_2017,
place={Cambridge}, 
edition={2}, 
title={Quantum Information Theory},
publisher={Cambridge University Press},
author={Wilde, Mark M.}, 
year={2017}
}

@article{Cech2026,
  title = {Revealing emergent many-body phenomena by analyzing large-scale space-time records of monitored quantum systems},
  author = {Cech, Marcel and De Fazio, Cecilia and Cea, Mar{\'i}a and Ba{\~n}uls, Mari Carmen and Lesanovsky, Igor and Carollo, Federico},
  journal = {Phys. Rev. A},
  volume = {113},
  issue = {6},
  pages = {L060201},
  year = {2026},
  month = {Jun},
  publisher = {American Physical Society},
  doi = {10.1103/flpc-1sls},
  url = {https://doi.org/10.1103/flpc-1sls}
}

@article{Cech2025,
  title = {Space-Time Correlations in Monitored Kinetically Constrained Discrete-Time Quantum Dynamics},
  author = {Cech, Marcel and Cea, Mar{\'i}a and Ba{\~n}uls, Mari Carmen and Lesanovsky, Igor and Carollo, Federico},
  journal = {Phys. Rev. Lett.},
  volume = {134},
  issue = {23},
  pages = {230403},
  year = {2025},
  month = {Jun},
  publisher = {American Physical Society},
  doi = {10.1103/2lxs-wccj},
  url = {https://doi.org/10.1103/2lxs-wccj}
}

@article{PhysRevA.111.022406,
  title = {Channel-based framework for phase estimation of multiple eigenvalues},
  author = {Jin, Yuan-De and Zhang, Shi-Yu and Ma, Wen-Long},
  journal = {Phys. Rev. A},
  volume = {111},
  issue = {2},
  pages = {022406},
  numpages = {17},
  year = {2025},
  month = {Feb},
  publisher = {American Physical Society},
  doi = {10.1103/PhysRevA.111.022406},
  url = {https://link.aps.org/doi/10.1103/PhysRevA.111.022406}
}

@article{PhysRevLett.125.080501,
  title = {Approaching Quantum-Limited Metrology with Imperfect Detectors by Using Weak-Value Amplification},
  author = {Xu, Liang and Liu, Zexuan and Datta, Animesh and Knee, George C. and Lundeen, Jeff S. and Lu, Yan-qing and Zhang, Lijian},
  journal = {Phys. Rev. Lett.},
  volume = {125},
  issue = {8},
  pages = {080501},
  numpages = {6},
  year = {2020},
  month = {Aug},
  publisher = {American Physical Society},
  doi = {10.1103/PhysRevLett.125.080501},
  url = {https://link.aps.org/doi/10.1103/PhysRevLett.125.080501}
}

@article{Coto2017,
  author    = {Coto, Raul and Montenegro, Víctor and Eremeev, Vitalie and Mundarain, Douglas and Orszag, Miguel},
  title     = {The power of a control qubit in weak measurements},
  journal   = {Sci. Rep.},
  year      = {2017},
  volume    = {7},
  number    = {1},
  pages     = {6351},
  month     = {July},
  doi       = {10.1038/s41598-017-05370-5},
  url       = {https://doi.org/10.1038/s41598-017-05370-5},
  issn      = {2045-2322}
}

@article{PhysRevA.92.012119,
  title = {Weak values in continuous weak measurements of qubits},
  author = {Qin, Lupei and Liang, Pengfei and Li, Xin-Qi},
  journal = {Phys. Rev. A},
  volume = {92},
  issue = {1},
  pages = {012119},
  numpages = {9},
  year = {2015},
  month = {Jul},
  publisher = {American Physical Society},
  doi = {10.1103/PhysRevA.92.012119},
  url = {https://link.aps.org/doi/10.1103/PhysRevA.92.012119}
}

@article{PhysRevLett.105.010405,
  title = {Measuring Small Longitudinal Phase Shifts: Weak Measurements or Standard Interferometry?},
  author = {Brunner, Nicolas and Simon, Christoph},
  journal = {Phys. Rev. Lett.},
  volume = {105},
  issue = {1},
  pages = {010405},
  numpages = {4},
  year = {2010},
  month = {Jul},
  publisher = {American Physical Society},
  doi = {10.1103/PhysRevLett.105.010405},
  url = {https://link.aps.org/doi/10.1103/PhysRevLett.105.010405}
}

@article{PhysRevLett.102.173601,
  title = {Ultrasensitive Beam Deflection Measurement via Interferometric Weak Value Amplification},
  author = {Dixon, P. Ben and Starling, David J. and Jordan, Andrew N. and Howell, John C.},
  journal = {Phys. Rev. Lett.},
  volume = {102},
  issue = {17},
  pages = {173601},
  numpages = {4},
  year = {2009},
  month = {Apr},
  publisher = {American Physical Society},
  doi = {10.1103/PhysRevLett.102.173601},
  url = {https://link.aps.org/doi/10.1103/PhysRevLett.102.173601}
}

@article{PhysRevA.107.063706,
  title = {Sequential unsharp measurement of photon polarization},
  author = {Adam, Peter and Di\'osi, Lajos},
  journal = {Phys. Rev. A},
  volume = {107},
  issue = {6},
  pages = {063706},
  numpages = {8},
  year = {2023},
  month = {Jun},
  publisher = {American Physical Society},
  doi = {10.1103/PhysRevA.107.063706},
  url = {https://link.aps.org/doi/10.1103/PhysRevA.107.063706}
}

@article{PhysRevA.91.032116,
  title = {Weak values as interference phenomena},
  author = {Dressel, Justin},
  journal = {Phys. Rev. A},
  volume = {91},
  issue = {3},
  pages = {032116},
  numpages = {14},
  year = {2015},
  month = {Mar},
  publisher = {American Physical Society},
  doi = {10.1103/PhysRevA.91.032116},
  url = {https://link.aps.org/doi/10.1103/PhysRevA.91.032116}
}

@article{PhysRevLett.109.150407,
  title = {Observing Trajectories with Weak Measurements in Quantum Systems in the Semiclassical Regime},
  author = {Matzkin, A.},
  journal = {Phys. Rev. Lett.},
  volume = {109},
  issue = {15},
  pages = {150407},
  numpages = {5},
  year = {2012},
  month = {Oct},
  publisher = {American Physical Society},
  doi = {10.1103/PhysRevLett.109.150407},
  url = {https://link.aps.org/doi/10.1103/PhysRevLett.109.150407}
}

@article{PhysRevA.76.044103,
  title = {Complex weak values in quantum measurement},
  author = {Jozsa, Richard},
  journal = {Phys. Rev. A},
  volume = {76},
  issue = {4},
  pages = {044103},
  numpages = {3},
  year = {2007},
  month = {Oct},
  publisher = {American Physical Society},
  doi = {10.1103/PhysRevA.76.044103},
  url = {https://link.aps.org/doi/10.1103/PhysRevA.76.044103}
}

@article{RevModPhys.86.1203,
  title = {Quantum channels and memory effects},
  author = {Caruso, Filippo and Giovannetti, Vittorio and Lupo, Cosmo and Mancini, Stefano},
  journal = {Rev. Mod. Phys.},
  volume = {86},
  issue = {4},
  pages = {1203--1259},
  numpages = {57},
  year = {2014},
  month = {Dec},
  publisher = {American Physical Society},
  doi = {10.1103/RevModPhys.86.1203},
  url = {https://link.aps.org/doi/10.1103/RevModPhys.86.1203}
}

@article{Ofek2016Nature,
  author  = {Ofek, Nissim and Petrenko, Andrei and Heeres, Reinier and Reinhold, Philip and Leghtas, Zaki and Vlastakis, Brian and Liu, Yehan and Frunzio, Luigi and Girvin, S. M. and Jiang, L. and Mirrahimi, Mazyar and Devoret, M. H. and Schoelkopf, R. J.},
  title   = {Extending the lifetime of a quantum bit with error correction in superconducting circuits},
  journal = {Nature},
  volume  = {536},
  pages   = {441--445},
  year    = {2016},
  doi     = {10.1038/nature18949}
}

@article{Eickbusch2022,
  author  = {Eickbusch, A. and Sivak, V. and Ding, A. Z. and Elder, S. S. and Jha, S. R. and Venkatraman, J. and Royer, B. and Girvin, S. M. and Schoelkopf, R. J. and Devoret, M. H.},
  title   = {Fast universal control of an oscillator with weak dispersive coupling to a qubit},
  journal = {Nat. Phys.},
  volume  = {18},
  pages   = {1464},
  year    = {2022},
  doi     = {10.1038/s41567-022-01776-9}
}

@article{Heeres2015PRL,
  title     = {Cavity State Manipulation Using Photon-Number Selective
               Phase Gates},
  author    = {Heeres, Reinier W. and Vlastakis, Brian and Holland, Eric
               and Krastanov, Stefan and Albert, Victor V.
               and Frunzio, Luigi and Jiang, Liang
               and Schoelkopf, Robert J.},
  journal   = {Phys. Rev. Lett.},
  volume    = {115},
  issue     = {13},
  pages     = {137002},
  numpages  = {5},
  year      = {2015},
  month     = {Sep},
  publisher = {American Physical Society},
  doi       = {10.1103/PhysRevLett.115.137002},
  url       = {https://link.aps.org/doi/10.1103/PhysRevLett.115.137002}
}

@article{Krastanov2015,
  title     = {Universal control of an oscillator with dispersive
               coupling to a qubit},
  author    = {Krastanov, Stefan and Albert, Victor V. and Shen, Chao
               and Zou, Chang-Ling and Heeres, Reinier W.
               and Vlastakis, Brian and Schoelkopf, Robert J.
               and Jiang, Liang},
  journal   = {Phys. Rev. A},
  volume    = {92},
  issue     = {4},
  pages     = {040303(R)},
  numpages  = {5},
  year      = {2015},
  month     = {Oct},
  publisher = {American Physical Society},
  doi       = {10.1103/PhysRevA.92.040303},
  url       = {https://link.aps.org/doi/10.1103/PhysRevA.92.040303}
}

@article{sun2014tracking,
  title     = {Tracking photon jumps with repeated quantum non-demolition
               parity measurements},
  author    = {Sun, L. and Petrenko, A. and Leghtas, Z.
               and Vlastakis, B. and Kirchmair, G. and Sliwa, K. M.
               and Narla, A. and Hatridge, M. and Shankar, S.
               and Blumoff, J. and Frunzio, L. and Mirrahimi, M.
               and Devoret, M. H. and Schoelkopf, R. J.},
  journal   = {Nature},
  volume    = {511},
  number    = {7510},
  pages     = {444--448},
  year      = {2014},
  month     = {Jul},
  doi       = {10.1038/nature13436},
  url       = {https://doi.org/10.1038/nature13436}
}

@article{Johnson2010QND,
  title     = {Quantum non-demolition detection of single microwave
               photons in a circuit},
  author    = {Johnson, B. R. and Reed, M. D. and Houck, A. A.
               and Schuster, D. I. and Bishop, Lev S. and Ginossar, E.
               and Gambetta, J. M. and DiCarlo, L. and Frunzio, L.
               and Girvin, S. M. and Schoelkopf, R. J.},
  journal   = {Nature Physics},
  volume    = {6},
  number    = {9},
  pages     = {663--667},
  year      = {2010},
  month     = {Sep},
  doi       = {10.1038/nphys1710},
  url       = {https://doi.org/10.1038/nphys1710}
}

@article{science.1152697,
author = {Onur Hosten  and Paul Kwiat },
title = {Observation of the Spin Hall Effect of Light via Weak Measurements},
journal = {Science},
volume = {319},
number = {5864},
pages = {787-790},
year = {2008},
doi = {10.1126/science.1152697},
URL = {https://www.science.org/doi/abs/10.1126/science.1152697}
}

@article{PhysRevLett.126.220801,
  title = {Enhanced Weak-Value Amplification via Photon Recycling},
  author = {Krafczyk, Courtney and Jordan, Andrew N. and Goggin, Michael E. and Kwiat, Paul G.},
  journal = {Phys. Rev. Lett.},
  volume = {126},
  issue = {22},
  pages = {220801},
  numpages = {5},
  year = {2021},
  month = {Jun},
  publisher = {American Physical Society},
  doi = {10.1103/PhysRevLett.126.220801},
  url = {https://link.aps.org/doi/10.1103/PhysRevLett.126.220801}
}

@article{PhysRevLett.111.160401,
  title = {Past Quantum States of a Monitored System},
  author = {Gammelmark, S\o{}ren and Julsgaard, Brian and M\o{}lmer, Klaus},
  journal = {Phys. Rev. Lett.},
  volume = {111},
  issue = {16},
  pages = {160401},
  numpages = {5},
  year = {2013},
  month = {Oct},
  publisher = {American Physical Society},
  doi = {10.1103/PhysRevLett.111.160401},
  url = {https://link.aps.org/doi/10.1103/PhysRevLett.111.160401}
}

@article{PhysRevA.89.012121,
  title = {Pre- and postselected quantum states: Density matrices, tomography, and Kraus operators},
  author = {Silva, Ralph and Guryanova, Yelena and Brunner, Nicolas and Linden, Noah and Short, Anthony J. and Popescu, Sandu},
  journal = {Phys. Rev. A},
  volume = {89},
  issue = {1},
  pages = {012121},
  numpages = {8},
  year = {2014},
  month = {Jan},
  publisher = {American Physical Society},
  doi = {10.1103/PhysRevA.89.012121},
  url = {https://link.aps.org/doi/10.1103/PhysRevA.89.012121}
}

@article{PhysRevA.96.062110,
  title = {Past observable dynamics of a continuously monitored qubit},
  author = {Garc\'{\i}a-Pintos, Luis Pedro and Dressel, Justin},
  journal = {Phys. Rev. A},
  volume = {96},
  issue = {6},
  pages = {062110},
  numpages = {10},
  year = {2017},
  month = {Dec},
  publisher = {American Physical Society},
  doi = {10.1103/PhysRevA.96.062110},
  url = {https://link.aps.org/doi/10.1103/PhysRevA.96.062110}
}

@article{kfn4-1fx4,
  title = {Retrodiction of measurement outcomes on a single quantum system revealing entanglement with its environment},
  author = {Pinske, Julien and M\o{}lmer, Klaus},
  journal = {Phys. Rev. A},
  volume = {112},
  issue = {3},
  pages = {032208},
  numpages = {12},
  year = {2025},
  month = {Sep},
  publisher = {American Physical Society},
  doi = {10.1103/kfn4-1fx4},
  url = {https://link.aps.org/doi/10.1103/kfn4-1fx4}
}

@article{Kim2018,
  author    = {Kim, Yosep and Kim, Yong-Su and Lee, Sang-Yun and Han, Sang-Wook and Moon, Sung and Kim, Yoon-Ho and Cho, Young-Wook},
  title     = {Direct quantum process tomography via measuring sequential weak values of incompatible observables},
  journal   = {Nat. Commun.},
  year      = {2018},
  month     = {jan},
  volume    = {9},
  number    = {1},
  pages     = {192},
  doi       = {10.1038/s41467-017-02511-2},
  url       = {https://doi.org/10.1038/s41467-017-02511-2},
  issn      = {2041-1723}
}

@article{Pfender2019NatCommun,
  author  = {Pfender, Matthias and Wang, Ping and Sumiya, Hitoshi and Onoda, Shinobu and Yang, Wen and Dasari, Durga Bhaktavatsala Rao and Neumann, Philipp and Pan, Xin-Yu and Isoya, Junichi and Liu, Ren-Bao and Wrachtrup, J{\"o}rg},
  title   = {High-resolution spectroscopy of single nuclear spins via sequential weak measurements},
  journal = {Nat. Commun.},
  volume  = {10},
  pages   = {594},
  year    = {2019},
  doi     = {10.1038/s41467-019-08544-z}
}

@article{Cujia2019Tracking,
  author  = {Cujia, K. S. and Boss, J. M. and Herb, K. and Zopes, J. and Degen, C. L.},
  title   = {Tracking the precession of single nuclear spins by weak measurements},
  journal = {Nature},
  volume  = {571},
  pages   = {230--233},
  year    = {2019},
  doi     = {10.1038/s41586-019-1334-9}
}

@article{pan2020weak,
  title={Weak-to-strong transition of quantum measurement in a trapped-ion system},
  author={Pan, Yiming and Zhang, Jie and Cohen, Eliahu and Wu, Chun-wang and Chen, Ping-Xing and Davidson, Nir},
  journal={Nat. Phys.},
  volume={16},
  number={12},
  pages={1206--1210},
  year={2020},
  publisher={Nature Publishing Group UK London},
  doi = {10.1038/s41567-020-0973-y},
  url = {https://doi.org/10.1038/s41567-020-0973-y}
}

@article{Groen2013,
  author  = {Groen, J. P. and Rist{\`e}, D. and Tornberg, L. and Cramer, J.
             and de Groot, P. C. and Picot, T. and Johansson, G. and DiCarlo, L.},
  title   = {Partial-Measurement Backaction and Nonclassical Weak Values
             in a Superconducting Circuit},
  journal = {Phys. Rev. Lett.},
  volume  = {111},
  pages   = {090506},
  year    = {2013},
  doi     = {10.1103/PhysRevLett.111.090506}
}

@article{Wang2019SingleMode,
  author = {Wang, W. and Wu, Y. and Ma, Y. and Cai, W. and Hu, L. and Mu, X. and Xu, Y. and Chen, Zi-Jie and Wang, H. and Song, Y. P. and Yuan, H. and Zou, C.-L. and Duan, L.-M. and Sun, L.},
  title = {Heisenberg-limited single-mode quantum metrology in a superconducting circuit},
  journal = {Nat. Commun.},
  volume = {10},
  pages = {4382},
  year = {2019},
  doi = {10.1038/s41467-019-12290-7}
}

@article{Wiseman2002Trajectories,
  author = {Wiseman, H. M.},
  title = {Weak values, quantum trajectories, and the cavity-{QED} experiment on wave-particle correlation},
  journal = {Phys. Rev. A},
  volume = {65},
  pages = {032111},
  year = {2002},
  doi = {10.1103/PhysRevA.65.032111}
}

@article{FranquetNazarov2017,
  author = {Franquet, A. and Nazarov, Yuli V.},
  title = {Probability distributions of continuous measurement results for conditioned quantum evolution},
  journal = {Phys. Rev. B},
  volume = {95},
  pages = {085427},
  year = {2017},
  doi = {10.1103/PhysRevB.95.085427}
}

\end{document}


\title{Supplemental Material for ``Weak Values in Postselected Quantum Trajectories"}
\author{Janarbek Yuanbek}
\thanks{These authors contributed equally to this work.}
\author{Yuan-De Jin}
\thanks{These authors contributed equally to this work.}
\author{Wen-Long Ma}
\email{Corresponding author: wenlongma@semi.ac.cn}

\affiliation{State Key Laboratory of Semiconductor Physics and Chip Technologies, Institute of Semiconductors, Chinese Academy of Sciences, Beijing 100083, China}
\affiliation{Center of Materials Science and Opto-Electronic Technology, University of Chinese Academy of Sciences, Beijing 100049, China}
\noaffiliation
\date{\today}
\begin{abstract}
This Supplemental Material derives the weak-value response of sequential measurement records and its realization through an ancilla-mediated Ramsey sequence. We first obtain the general conditional response and its reduction to a symmetric binary instrument. We then present the qubit and oscillator examples, including their exact finite-strength record statistics, weak-response limits, probability landscapes, and conditional displacement-to-width ratios.
\end{abstract}

\maketitle
\tableofcontents

\section{Derivation of the general weak-instrument response}

We relate the weak value to the statistics of a complete sequential weak-measurement record. Consider a sequence of $m$ weak measurements with outcomes $\boldsymbol{\alpha}=(\alpha_1,\ldots,\alpha_m)$~\cite{PhysRevA.107.012217}. At step $k$, the outcome-dependent Kraus operator admits the weak-response expansion~\cite{PhysRevLett.95.110409}
\begin{equation}
M_{\alpha_k}
=
\sqrt{p_{\alpha_k}}
\left[I+\epsilon L_{\alpha_k}\right]
+O(\epsilon^2),
\label{eq:single-step}
\end{equation}
where $\epsilon\ll1$ is the weak-response parameter, $p_{\alpha_k}$ is the zeroth-order outcome probability, and $L_{\alpha_k}$ is the corresponding response operator. To first order, $M_{\alpha_k}^{\dagger}M_{\alpha_k}=p_{\alpha_k}[I+\epsilon(L_{\alpha_k}+L_{\alpha_k}^{\dagger})]+O(\epsilon^2)$. Therefore, matching the zeroth- and first-order terms in the completeness condition $\sum_{\alpha_k}M_{\alpha_k}^{\dagger}M_{\alpha_k}=I+O(\epsilon^2)$ gives
$
\sum_{\alpha_k}p_{\alpha_k}=1,
$
and
$
\sum_{\alpha_k}p_{\alpha_k}
\left(L_{\alpha_k}+L_{\alpha_k}^{\dagger}\right)=0.
$
We restrict to the commuting measurement channel considered in the main text, for which all Kraus operators are functions of the same Hermitian observable. The complete-record Kraus operator is
$K_{\boldsymbol{\alpha}}=M_{\alpha_m}\cdots M_{\alpha_1}$. Introducing the zeroth-order record probability
$p_0(\boldsymbol{\alpha})=\prod_{k=1}^{m}p_{\alpha_k}$, substitution of Eq.~\eqref{eq:single-step} gives
\begin{align}
K_{\boldsymbol{\alpha}}
&=
\sqrt{p_0(\boldsymbol{\alpha})}
\prod_{k=1}^{m}\left[I+\epsilon L_{\alpha_k}\right]
+O(\epsilon^2)
\approx
\sqrt{p_0(\boldsymbol{\alpha})}
(I+\epsilon\sum_{k=1}^{m}L_{\alpha_k}).
\end{align}

Thus, defining the accumulated response operator
$
\mathcal L_{\boldsymbol{\alpha}}
=
\sum_{k=1}^{m}L_{\alpha_k},
$
the full-record operator becomes
\begin{equation}
K_{\boldsymbol{\alpha}}
\approx
\sqrt{p_0(\boldsymbol{\alpha})}
\left[I+\epsilon\mathcal L_{\boldsymbol{\alpha}}\right].
\label{eq:record-kraus}
\end{equation}

The first-order product rule can be seen already from two factors: $(I+\epsilon L_2)(I+\epsilon L_1)=I+\epsilon(L_1+L_2)+\epsilon^2L_2L_1$. Hence, all ordering-dependent products of the response operators first appear at $O(\epsilon^2)$ and do not contribute to the leading weak response. Consequently, to first order, the effect of the complete measurement trajectory is determined by the additive accumulated response $\mathcal L_{\boldsymbol{\alpha}}$, rather than by the detailed ordering of individual weak steps~\cite{PhysRevA.109.022224,PhysRevLett.134.080802}.

%
Let the initial target state be $\rho_i$, and let the final postselection be described by $\Pi_f$. For a nonzero initial overlap $P_f^{(0)}=\operatorname{Tr}(\Pi_f\rho_i)>0$, we use
\begin{equation}
B_w=\frac{\operatorname{Tr}(\Pi_f B\rho_i)}{P_f^{(0)}}
\label{add:general-weak-value}
\end{equation}
for any target operator $B$. 
This notation applies to both the measured observable $\mathcal A$ and the accumulated response $\mathcal L_{\boldsymbol\alpha}$. For pure pre- and postselected states, it reduces to the Aharonov--Albert--Vaidman weak value $B_w=\langle\psi_f|B|\psi_i\rangle/\langle\psi_f|\psi_i\rangle$~\cite{PhysRevLett.60.1351}.
The joint probability of observing $\boldsymbol\alpha$ and subsequently obtaining the postselection outcome is $P_{\rm post}(\boldsymbol\alpha)=\operatorname{Tr}[\Pi_f K_{\boldsymbol\alpha}\rho_i K_{\boldsymbol\alpha}^{\dagger}]$~\cite{PhysRevA.81.012103}.

Using Eq.~\eqref{eq:record-kraus}, the density operator appearing in this trace is
\begin{align}
K_{\boldsymbol{\alpha}}\rho_iK_{\boldsymbol{\alpha}}^{\dagger} &\approx p_0(\boldsymbol{\alpha}) \left[I+\epsilon\mathcal L_{\boldsymbol{\alpha}}\right] \rho_i \left[I+\epsilon\mathcal L_{\boldsymbol{\alpha}}^{\dagger}\right] \approx p_0(\boldsymbol{\alpha}) \left[ \rho_i +\epsilon\mathcal L_{\boldsymbol{\alpha}}\rho_i +\epsilon\rho_i\mathcal L_{\boldsymbol{\alpha}}^{\dagger} \right],
\end{align}
consequently,
\begin{equation}
P_{\rm post}(\boldsymbol{\alpha})
\approx
p_0(\boldsymbol{\alpha})
\left[
\operatorname{Tr}(\Pi_f\rho_i)
+\epsilon\operatorname{Tr}(\Pi_f\mathcal L_{\boldsymbol{\alpha}}\rho_i)
+\epsilon\operatorname{Tr}(\Pi_f\rho_i\mathcal L_{\boldsymbol{\alpha}}^{\dagger})
\right].
\label{eq:joint-expanded}
\end{equation}
%

The two first-order trace terms are complex conjugates, since $\rho_i$ and $\Pi_f$ are Hermitian and the trace is cyclic. Using the weak-value notation introduced above, Eq.~\eqref{eq:joint-expanded} therefore becomes

\begin{equation}
P_{\rm post}(\boldsymbol{\alpha})
\approx
p_0(\boldsymbol{\alpha})P_f^{(0)}
(
1+2\epsilon\operatorname{Re}[(\mathcal L_{\boldsymbol{\alpha}})_w]
).
\label{eq:joint-weak}
\end{equation}

The normalized distribution conditioned on successful postselection is
$P_f(\boldsymbol{\alpha})=P_{\rm post}(\boldsymbol{\alpha})/\sum_{\boldsymbol{\alpha}}P_{\rm post}(\boldsymbol{\alpha})$. In summing Eq.~\eqref{eq:joint-weak}, the zeroth-order term gives $\sum_{\boldsymbol{\alpha}}p_0(\boldsymbol{\alpha})=1$, while the first-order term gives the $p_0$-weighted record average. The denominator is therefore
\begin{align}
\sum_{\boldsymbol{\alpha}}P_{\rm post}(\boldsymbol{\alpha})
&\approx
P_f^{(0)}
(
1+2\epsilon\operatorname{Re}[
\left\langle(\mathcal L_{\boldsymbol{\alpha}})_w\right\rangle_0]
),
\end{align}
where $\langle F\rangle_0=\sum_{\boldsymbol{\alpha}}p_0(\boldsymbol{\alpha})F(\boldsymbol{\alpha})$. Expanding the inverse denominator as $(1+x)^{-1}\approx1-x$ and retaining only $O(\epsilon)$ terms yields
\begin{equation}
P_f(\boldsymbol{\alpha})
\approx
p_0(\boldsymbol{\alpha})
(
1+2\epsilon\operatorname{Re}
[
(\mathcal L_{\boldsymbol{\alpha}})_w
-
\left\langle(\mathcal L_{\boldsymbol{\alpha}})_w\right\rangle_0
]
).
\label{eq:conditional-general}
\end{equation}

The subtraction term is required for normalization and removes the record-independent part of the postselection response. Thus, only the record-dependent variation of the weak response can modify the shape or moments of the conditional distribution.

For an arbitrary real record observable $X(\boldsymbol{\alpha})$, its conditional average is
$\langle X\rangle_f=\sum_{\boldsymbol{\alpha}}X(\boldsymbol{\alpha})P_f(\boldsymbol{\alpha})$. Inserting Eq.~\eqref{eq:conditional-general} gives
\begin{align}
\langle X\rangle_f &\approx \langle X\rangle_0 +2\epsilon \sum_{\boldsymbol{\alpha}} p_0(\boldsymbol{\alpha})X(\boldsymbol{\alpha}) \operatorname{Re} \left[ (\mathcal L_{\boldsymbol{\alpha}})_w - \left\langle(\mathcal L_{\boldsymbol{\alpha}})_w\right\rangle_0 \right] = \langle X\rangle_0 +2\epsilon\operatorname{Re} \left[ \left\langle X(\boldsymbol{\alpha})(\mathcal L_{\boldsymbol{\alpha}})_w\right\rangle_0 - \langle X\rangle_0 \left\langle(\mathcal L_{\boldsymbol{\alpha}})_w\right\rangle_0 \right].
\label{eq.final}
\end{align}

%
%
Equation~\eqref{eq:conditional-general} gives the normalized conditional record distribution, whereas Eq.~\eqref{eq.final} gives the corresponding leading response of any real record observable $X(\boldsymbol\alpha)$.

\section{Reduction to the symmetric binary weak-measurement model}

For the symmetric binary instrument,
\begin{equation}
M_{\alpha_k}=\frac{1}{\sqrt2}\left[I+\frac{\alpha_k\theta}{2}\mathcal A\right]+O(\theta^2),
\qquad\alpha_k=\pm1.
\label{eq:binary-step}
\end{equation}
Defining $S=\sum_{k=1}^{m}\alpha_k$ and $X=S/(2m)=q/m-1/2$, where $q$ counts the positive outcomes, gives $p_0(\boldsymbol\alpha)=2^{-m}$, $\epsilon=\theta$, and $\mathcal L_{\boldsymbol\alpha}=S\mathcal A/2$. 
Hence $(\mathcal L_{\boldsymbol\alpha})_w=S\mathcal A_w/2$. Since $\langle S\rangle_0=0$, Eq.~\eqref{eq:conditional-general} reduces to
\begin{equation}
P_f(\boldsymbol\alpha)\simeq2^{-m}\left[1+\theta S\operatorname{Re}(\mathcal A_w)\right].
\label{eq:binary-record-probability}
\end{equation}
The balanced zeroth-order ensemble has $\langle S^2\rangle_0=m$. Thus the leading conditional and unconditioned means are
\begin{equation}
\langle X\rangle_f\simeq\frac{\theta}{2}\operatorname{Re}(\mathcal A_w),
\qquad
\langle X\rangle_{\rm no}\simeq\frac{\theta}{2}\langle\mathcal A\rangle_i,
\label{eq:binary-means}
\end{equation}
where $\langle\mathcal A\rangle_i=\operatorname{Tr}(\mathcal A\rho_i)$. The displacement relative to the unconditioned mean is therefore
\begin{equation}
\Delta\langle X\rangle\simeq\frac{\theta}{2}\left[\operatorname{Re}(\mathcal A_w)-\langle\mathcal A\rangle_i\right].
\label{eq:binary-displacement}
\end{equation}
These relations do not require $\mathcal A^2=I$ and apply to both models within their controlled weak-response regimes.

For typical balanced records, $S=O(\sqrt m)$, so the conditional record expansion requires $\sqrt m\,|\theta\mathcal A_w|\ll1$ together with perturbative accumulated backaction. At fixed $m$ on finite support, the next record-amplitude term is
\begin{equation*}
K_{\boldsymbol\alpha}=2^{-m/2}\left[
I+\frac{\theta S}{2}\mathcal A
+\theta^2\left(\frac{S^2}{8}-\frac m4\right)\mathcal A^2+\cdots\right].
\end{equation*}
%
%
For fixed $m$, finite support, and nonzero $P_f^{(0)}$, the success probability satisfies
\begin{equation*}
\frac{P_f(m,\theta)}{P_f^{(0)}}=
1+\frac{m\theta^2}{4}\left[
\frac{\operatorname{Tr}(\Pi_f\mathcal A\rho_i\mathcal A)}{P_f^{(0)}}
-\operatorname{Re}(\mathcal A^2)_w\right]+O_m(\theta^4).
\end{equation*}
For pure preselection and rank-one postselection, the first term in brackets reduces to $|\mathcal A_w|^2$. The remainder need not be uniform as $m$ increases or $P_f^{(0)}$ approaches zero.

The same balanced statistics give $\operatorname{Var}_0(S)=\langle S^2\rangle_0-\langle S\rangle_0^2=m$, and therefore $\operatorname{Var}_0(X)=\operatorname{Var}_0(S)/(4m^2)=1/(4m)$. Thus, the unnormalized sum $S$ acquires a bias proportional to $m\theta\operatorname{Re}[\mathcal{A}_w]$, whereas the normalized polarization has an $m$-independent leading shift. Repeated measurements instead narrow the distribution of $X$ as $m^{-1/2}$, thereby increasing the conditional distinguishability of the weak-value-induced displacement.

\section{Ramsey implementation of the binary weak measurement}

We use the time-independent interaction and Ramsey convention of the main text~\cite{66yg-qs8x}. With $H_{\rm int}=g\sigma_z^q\otimes\mathcal A$, $\theta=2gt$, and $\hbar=1$, the propagator is
\begin{equation}
U=|0\rangle_q\langle0|\otimes e^{-i\theta\mathcal A/2}
+|1\rangle_q\langle1|\otimes e^{i\theta\mathcal A/2}.
\label{eq:conditional-propagator}
\end{equation}
The ancilla is initialized in $|0\rangle_q$, acted on by the two $\pi/2$ rotations $R_{\varphi_1}$ and $R_{\varphi_2}$, and measured in the computational basis. Assigning $\alpha=+1$ to $|0\rangle_q$ and $\alpha=-1$ to $|1\rangle_q$, and writing $\Delta\varphi=\varphi_1-\varphi_2$, gives, up to outcome-dependent global phases,
\begin{equation}
M_\alpha=\frac12\left(e^{-i\theta\mathcal A/2}
-\alpha e^{i\Delta\varphi}e^{i\theta\mathcal A/2}\right).
\label{eq:ramsey-phase-instrument}
\end{equation}
At the balanced point $\Delta\varphi=\pi/2$, another choice of these global phases yields
\begin{equation}
M_\alpha=\frac1{\sqrt2}\left[
\cos\left(\frac{\theta\mathcal A}{2}\right)
+\alpha\sin\left(\frac{\theta\mathcal A}{2}\right)\right].
\label{add:main-ramsey}
\end{equation}
Its first-order expansion is precisely the binary instrument in Eq.~\eqref{eq:binary-step}. For pure pre- and postselected states with fixed nonzero overlap, the conditional single-step probability is
\begin{equation}
P_f(\alpha)=
\frac{|\langle\psi_f|M_\alpha|\psi_i\rangle|^2}
{\sum_{\beta=\pm1}|\langle\psi_f|M_\beta|\psi_i\rangle|^2}
=\frac12\left[1+\alpha\theta\operatorname{Re}(\mathcal A_w)\right]+O(\theta^2).
\label{eq:ramsey-single-probability}
\end{equation}
The complete-sequence response then follows from Sec.~S2, with target postselection performed only after all $m$ measurements.

\section{Detailed calculation for the qubit sequential weak-measurement model}

%
%
%
We now specialize the binary instrument to $\mathcal A=\sigma_z$, retaining the weak-value and record notation introduced above. The following expressions describe the exact finite-strength statistics for the two postselection settings $j=1,2$. Since $\mathcal A^2=I$, the Kraus operator becomes

\begin{equation}
M_\alpha
=
\frac{1}{\sqrt{2}}
\left[
\cos\left(\frac{\theta}{2}\right)I
+\alpha\sin\left(\frac{\theta}{2}\right)\mathcal{A}
\right],
\qquad \alpha=\pm1,
\end{equation}
where $\theta=2gt$. Since $\mathcal{A}=\mathcal{A}^\dagger$ and $\mathcal{A}^2=I$, each $M_\alpha$ is Hermitian and
\begin{align}
M_\alpha^\dagger M_\alpha &= \frac12 \left[ \cos^2\frac\theta2+\sin^2\frac\theta2 +2\alpha\cos\frac\theta2\sin\frac\theta2 \mathcal{A} \right] = \frac12\left[I+\alpha\sin\theta\,\mathcal{A}\right].
\end{align}

Summing over $\alpha=\pm1$ cancels the term odd in $\alpha$, while the two $I/2$ terms add to $I$, giving $\sum_\alpha M_\alpha^\dagger M_\alpha=I$. The single-step outcome probability follows from $p_\alpha=\operatorname{Tr}(\rho_iM_\alpha^\dagger M_\alpha)$ and $\operatorname{Tr}(\rho_i)=1$, and is therefore
\begin{equation}
p_\alpha
=
\operatorname{Tr}(M_\alpha\rho_iM_\alpha^\dagger)
=
\frac12\left[1+\alpha\sin\theta\,\langle \mathcal{A}\rangle_i\right]
\approx
\frac12\left[1+\alpha\theta\langle \mathcal{A}\rangle_i\right].
\end{equation}

For $|\psi_i\rangle=(|0\rangle+|1\rangle)/\sqrt2$ and $\mathcal{A}=\sigma_z$, $\langle \mathcal{A}\rangle_i=0$, so the no-postselection record is unbiased.
For this balanced input, the two final states in the main text are
\begin{align}
|\psi_f^1\rangle
&=\cos\frac{\theta_f}{2}|0\rangle
+e^{i\phi_f}\sin\frac{\theta_f}{2}|1\rangle,
\label{add:qubit-state-one}\\
|\psi_f^2\rangle
&=\frac{|0\rangle-\sqrt3|1\rangle}{2}.
\label{add:qubit-state-two}
\end{align}
Each state defines a separate setting with effect $\Pi_f^j=|\psi_f^j\rangle\langle\psi_f^j|$, where $j=1,2$.
Postselection is performed once, after the complete sequence of $m$ weak measurements.
Equation~\eqref{add:general-weak-value} gives
\begin{align}
\mathcal{A}_w^1 &=\frac{\cos(\theta_f/2)-e^{-i\phi_f}\sin(\theta_f/2)} {\cos(\theta_f/2)+e^{-i\phi_f}\sin(\theta_f/2)} =\frac{\cos\theta_f+i\sin\theta_f\sin\phi_f} {1+\sin\theta_f\cos\phi_f}.
\label{add:qubit-weak-one}
\end{align}
For the second state, the corresponding result is $\mathcal{A}_w^2=-(2+\sqrt3)$.
Their initial overlap probabilities are $P_{f,1}^{(0)}=(1+\sin\theta_f\cos\phi_f)/2$ and $P_{f,2}^{(0)}=(2-\sqrt3)/4$.
The first state has Bloch vector $(\sin\theta_f\cos\phi_f,\sin\theta_f\sin\phi_f,\cos\theta_f)$, while the second has $(-\sqrt3/2,0,-1/2)$.
These results are the two-state specializations of the main-text Bloch-vector weak value.

We use the collective variables $S$ and $X$ defined in Sec.~S2. 
Because every $M_\alpha$ is a function of $I$ and $\mathcal{A}$, the Kraus operators commute, and the product depends only on $q$. Setting $c=\cos(\theta/2)$ and $d=\sin(\theta/2)$ gives
\begin{equation}
%
K(q)
=
2^{-m/2}(cI+d\mathcal{A})^q(cI-d\mathcal{A})^{m-q}.
\label{eq:exact-record-product}
\end{equation}

Introduce the projectors onto the eigenspaces of $\mathcal{A}$,
$\Pi_\pm=(I\pm \mathcal{A})/2$. Since $I=\Pi_++\Pi_-$ and $\mathcal{A}=\Pi_+-\Pi_-$, one has $cI+d\mathcal{A}=(c+d)\Pi_++(c-d)\Pi_-$ and the corresponding expression with $d\to-d$. Equivalently, using $\mathcal{A}\Pi_\pm=\pm\Pi_\pm$,
\begin{equation}
cI\pm d\mathcal{A}
=(c\pm d)\Pi_++(c\mp d)\Pi_-.
\end{equation}

Using $\Pi_+\Pi_-=0$ and $\Pi_\pm^2=\Pi_\pm$, Eq.~\eqref{eq:exact-record-product} becomes
\begin{equation}
%
K(q)
=
2^{-m/2}
\left[
(c+d)^q(c-d)^{m-q}\Pi_+
+
(c-d)^q(c+d)^{m-q}\Pi_-
\right].
\end{equation}

This is the exact record of the Kraus operator for the dichotomic model. It also shows explicitly why the full trajectory can be grouped only by the number $q$ of positive outcomes: for commuting measurements of a fixed dichotomic observable, no additional ordering information is required.
The exact normalized distribution for either final state can be written directly in the main-text record notation,
\begin{equation}
P_{\rm ps}^j(q)
=\binom mq\frac{|\langle\psi_f^j|K(q)|\psi_i\rangle|^2}
{\operatorname{Tr}[\Pi_f^j\Phi^m(\rho_i)]}.
\label{add:exact-grouped-record}
\end{equation}
Here $\Phi(\rho)=\sum_{\alpha=\pm1}M_\alpha\rho M_\alpha^\dagger$ is the measurement channel, and the binomial factor counts records with the same $q$.
Writing $\Pi_f^j=(I+\mathbf r^j\cdot\boldsymbol\sigma)/2$, the success probability is $\operatorname{Tr}[\Pi_f^j\Phi^m(\rho_i)]=[1+r_x^j(\cos\theta)^m]/2$.
The exact conditional mean is
\begin{equation}
\langle X\rangle_{\rm ps}^j
=\frac{r_z^j\sin\theta}{2[1+r_x^j(\cos\theta)^m]}.
\label{add:qubit-exact-mean}
\end{equation}
Thus the normalization includes accumulated measurement backaction and reduces to $P_{f,j}^{(0)}$ at zero strength.
Writing $D_m^j=1+r_x^j(\cos\theta)^m$, the exact conditional variance is
\begin{equation*}
\operatorname{Var}_{\rm ps}^j(X)=\frac{1}{4m}
+\frac{(m-1)\sin^2\theta}{4mD_m^j}
-\left(\frac{r_z^j\sin\theta}{2D_m^j}\right)^2.
\end{equation*}
For comparison, the unconditioned variance is
$\operatorname{Var}_{\rm no}(X)=\cos^2\theta/(4m)+\sin^2\theta/4$.
These formulas provide the finite-strength reference for the leading width.

The weak-response limit follows directly from Eq.~\eqref{eq:binary-record-probability}. Grouping the records by the number $q$ of positive outcomes gives
\begin{equation}
P_{\rm ps}^j(q)\simeq2^{-m}\binom mq
\left[1+\theta\operatorname{Re}(\mathcal A_w^j)(2q-m)\right].
\label{eq:qubit-grouped-weak}
\end{equation}
For the balanced qubit input, $\langle\mathcal A\rangle_i=0$, so Eq.~\eqref{eq:binary-displacement} gives $\Delta\langle X\rangle_j\simeq(\theta/2)\operatorname{Re}(\mathcal A_w^j)$. At finite strength, the distribution and its normalization are evaluated using Eqs.~\eqref{add:exact-grouped-record} and \eqref{add:qubit-exact-mean}.
For $\rho_i=(I+\sigma_x)/2$ and $\Pi_f=(I+\mathbf r\cdot\boldsymbol\sigma)/2$, the Pauli trace identities give $P_f^{(0)}=(1+r_x)/2$ and $\operatorname{Tr}(\Pi_f\sigma_z\rho_i)=(r_z+ir_y)/2$. Therefore,
\begin{equation}
\mathcal A_w(\mathbf r)=\frac{r_z+ir_y}{1+r_x}.
\end{equation}
This expression gives the weak values for the two specified qubit postselections.

The leading record bias can be represented by a Bernoulli surrogate with $p_{\rm ps}=1/2+\theta\operatorname{Re}(\mathcal A_w)/2$,
\begin{equation}
P_B(q)=\binom mq p_{\rm ps}^q(1-p_{\rm ps})^{m-q}.
\end{equation}
For $x=q/m=1/2+X$, its large-$m$ probability landscape is~\cite{394k-lj6s}
\begin{equation}
F_B(X)=-\ln P_B(q)
=m\left[x\ln\frac{x}{p_{\rm ps}}+(1-x)\ln\frac{1-x}{1-p_{\rm ps}}\right]+O(\ln m).
\end{equation}
Its mode satisfies $X_{\rm peak}^{B}=p_{\rm ps}-1/2+O(1/m)$, recovering the leading mean polarization. The surrogate matches the sequential statistics only through first order in $\theta$; finite-strength peaks and widths must be obtained from the exact conditional distribution.

The symmetry of the zeroth-order record ensemble gives $\langle S^3\rangle_0=0$, so the leading conditional width is
\begin{equation}
\sigma_X=\sqrt{\operatorname{Var}_{\rm ps}(X)}\simeq\frac{1}{2\sqrt m}.
\end{equation}
Combining this width with Eq.~\eqref{eq:binary-displacement} gives
\begin{equation}
\mathcal R_{\rm ps}^{\rm cond}
=\frac{|\langle X\rangle_{\rm ps}-\langle X\rangle_{\rm no}|}{\sigma_X}
\simeq\theta|\operatorname{Re}(\mathcal A_w)-\langle\mathcal A\rangle_i|\sqrt m.
\label{add:main-text-snr}
\end{equation}
For the balanced qubit, $\langle\mathcal A\rangle_i=0$. Second-order width corrections require the full sequential dynamics.

\begin{figure}[t]
\centering
\includegraphics[width=\linewidth]{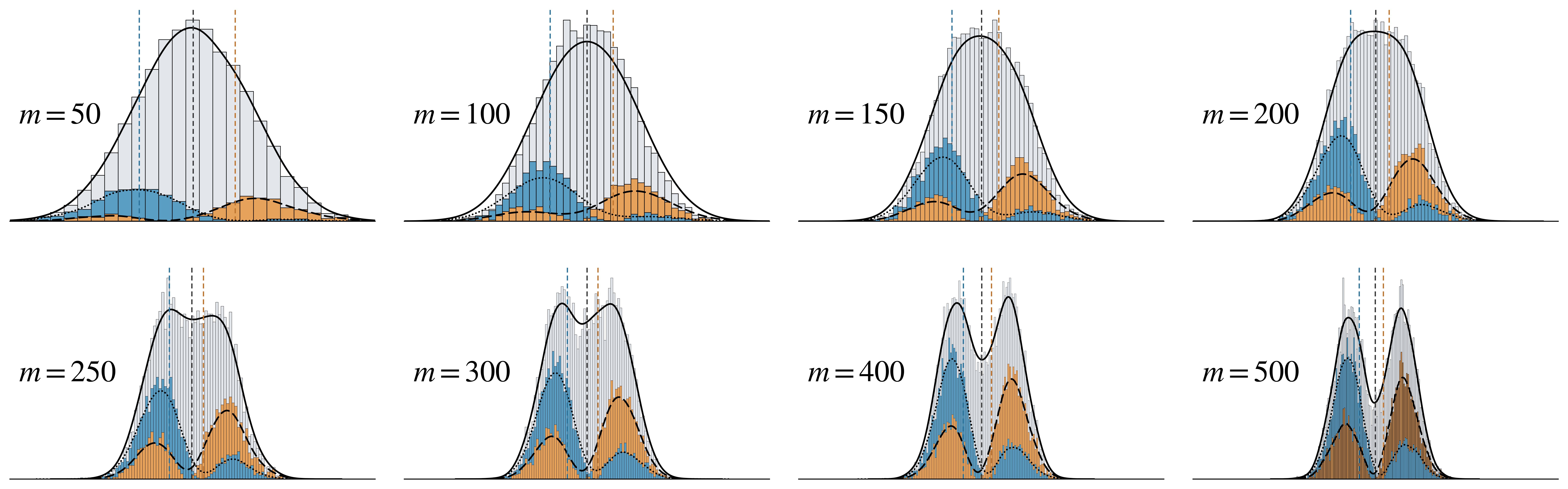} 
\caption{
Sequential weak-measurement record statistics for $\mathcal{A}=\sigma_z$ and $|\psi_i\rangle=(|0\rangle+|1\rangle)/\sqrt2$.
The upper and lower rows show $m=50,100,150,200$ and $m=250,300,400,500$, respectively, from left to right.
Gray, orange, and blue bars correspond to no postselection, $|\psi_f^1\rangle$, and $|\psi_f^2\rangle$.
The final states and their analytic weak values are given above in Sec.~S4.
The first setting uses $\theta_f=5\pi/12$ and $\phi_f=\pi$.
Each setting uses $10^4$ initial trajectories per $m$, with final postselection after the last measurement.
Here $g=1$, $t=\pi/80$, $\theta=2gt=\pi/40$, and $\Delta\varphi=\pi/2$.
The horizontal variable is $X$; bar heights are record counts divided by $m$, with bin width $1/m$.
Black solid, dashed, and dotted curves show the respective smoothed histograms, with Gaussian width one bin for $m=50$ and $3.5$ bins otherwise.
Colored vertical dashed lines mark sample means.
Equation~\eqref{add:exact-grouped-record} gives the normalized conditional probabilities underlying these scaled counts. The displayed count scale is not a normalized conditional probability, and the black curves are smoothed data rather than independent theoretical predictions.
\label{SI-Fig.1} }
\end{figure}

Within the joint weak-response window, repeated weak measurements leave the leading normalized displacement independent of $m$, while narrowing the conditional distribution. This scaling requires perturbative accumulated backaction; sufficient conditions for typical balanced qubit records are $m\theta^2\ll1$ and $\sqrt m\,|\theta\mathcal A_w^j|\ll1$. Beyond that regime, Eq.~\eqref{add:exact-grouped-record} gives the finite-strength distribution shown in Fig.~\ref{SI-Fig.1}.

For $N_{\rm raw}$ initial trajectories, the expected retained number is $\operatorname{E}[N_{\rm succ}^j]=N_{\rm raw}\operatorname{Tr}[\Pi_f^j\Phi^m(\rho_i)]$.
The standard error of the conditional sample mean is approximately $\sigma_X^j/\sqrt{N_{\rm succ}^j}$, which differs from the width of a single-record distribution. For independently estimated conditional and unconditioned means,
\begin{equation*}
\operatorname{SE}(\bar X_{\rm ps}^j-\bar X_{\rm no})
\simeq\sqrt{\frac{(\sigma_X^j)^2}{N_{\rm succ}^j}
+\frac{\sigma_{\rm no}^2}{N_{\rm no}}}.
\end{equation*}
Here, $\operatorname{SE}$ denotes the standard error of the difference between the two sample means. The quantities $\sigma_X^j$ and $\sigma_{\rm no}$ are the standard deviations of $X$ in the postselected and unconditioned ensembles, respectively, while $N_{\rm succ}^j$ and $N_{\rm no}$ are the actual numbers of trajectories used to estimate the corresponding means.
If the means use the same original trajectories, the covariance must be included, or whole trajectories and their acceptance indicators must be resampled together. Actual retained sample counts must be distinguished from their expectations.

\begin{figure}[b]
\centering
\includegraphics[width=\linewidth]{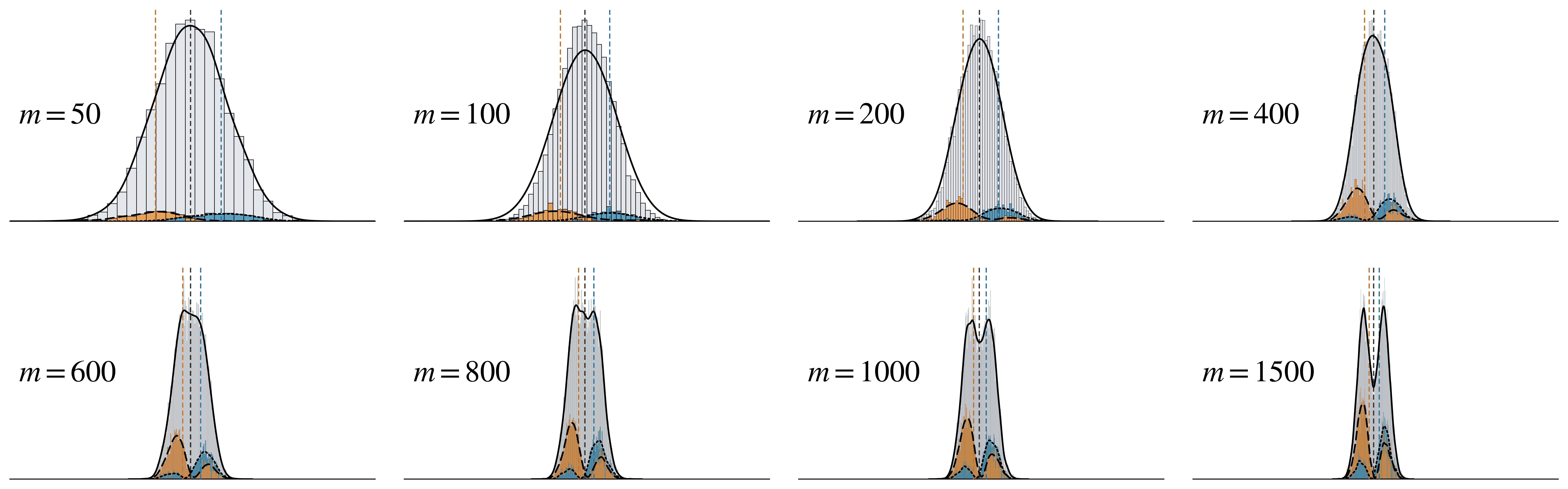} 
\caption{
Sequential weak-measurement record statistics for $\mathcal A=\hat a^\dagger\hat a$ and $|\psi_i\rangle=(|0\rangle+|4\rangle)/\sqrt2$.
The upper and lower rows show $m=50,100,200,400$ and $m=600,800,1000,1500$, respectively, from left to right.
Gray bars show records without postselection.
Orange bars correspond to $|\psi_f^1\rangle$ with $\theta_f=\vartheta_f=\pi/2$ and $\phi_f=\pi$.
Blue bars correspond to the even-cat state $|\psi_f^2\rangle$ with $\alpha=e^{i\pi/4}(5\sqrt{24}/3)^{1/4}$ and $\chi=0$.
Their analytic weak values are given in Eqs.~\eqref{add:osc-weak-one} and \eqref{add:osc-weak-two}.
Both acceptance tests use the same $10^4$ initial trajectories for each $m$.
Here $g=1$, $t=\pi/300$, $\theta=2gt=\pi/150$, and $\Delta\varphi=\pi/2$, with $50$ number states in the numerical Hilbert space.
Bars show raw counts in bins of width $1/m$ along $X=q/m-1/2$.
Black solid, dashed, and dotted curves show the corresponding smoothed histograms, with width one bin for $m=50$ and $3.5$ bins otherwise.
Colored vertical dashed lines mark sample means.
Equations~\eqref{add:exact-grouped-record}, \eqref{add:osc-exact-record}, and \eqref{add:osc-exact-success} give the exact conditional statistics. Raw count heights depend on the retained sample size; the black curves are smoothed data rather than independent theoretical predictions.
\label{SI-Fig.2} }
\end{figure}

\paragraph{Normalization and controlled  finite-sequence comparison.}
For a bin at $X=q/m-1/2$, empirical conditional probability mass is $C_q/N_{\rm succ}$; the unconditioned mass is $C_q/N_{\rm raw}$. For bin width $\Delta X=1/m$, the corresponding conditional density is $mC_q/N_{\rm succ}$. Probability masses sum to unity, whereas densities integrate to unity after multiplication by the bin width. The count scales in Figs.~\ref{SI-Fig.1} and \ref{SI-Fig.2} must therefore be distinguished from normalized conditional distributions.

For a deterministic test of the weak-response formulas, we use the initial and final states specified in Secs.~S4 and S5 and set $\theta=10^{-3}$ for all four postselection settings and $m=50,100,150,200,250,300,400,500$. 
Direct summation of the exact grouped probabilities gives maximum relative deviations $1.6201\%$ for the mean displacement, $1.5816\%$ for the leading width, and $0.0392\%$ for the conditional displacement-to-width ratio. Each deviation is normalized by the corresponding leading-order prediction. These are comparisons of exact probabilities with the weak-response formulas, not new Monte Carlo data.
At $m=50$, the exact displacements are $0.00379519$, $-0.00186572$, $-0.00581889$, and $0.00399696$ for the first and second qubit postselections and the first and cat oscillator postselections, respectively. The corresponding success probabilities are $0.0170492$, $0.0669981$, $0.0214820$, and $0.0254532$.
The original couplings used in Figs.~\ref{SI-Fig.1} and \ref{SI-Fig.2} instead illustrate finite-strength statistics outside the joint perturbative window.

\section{Detailed calculation for the oscillator model}
%
For $\mathcal A=\hat a^\dagger\hat a$, write $|\psi_i\rangle=\sum_n c_n|n\rangle$ and $|\psi_f\rangle=\sum_n d_n|n\rangle$. Equation~\eqref{add:general-weak-value} gives
\begin{equation}
\mathcal A_w=\frac{\displaystyle\sum_{n=0}^{\infty}n\,d_n^*c_n}
{\displaystyle\sum_{n=0}^{\infty}d_n^*c_n},
\label{add:osc-general-weak-value}
\end{equation}
assuming a nonzero denominator and convergence of the sums. The specified finite-support input reduces these sums to the populated number states.
For the vacuum--four-photon input, let $d_n^j=\langle n|\psi_f^j\rangle$. Then $\mathcal A_w^j=4d_4^{j*}/(d_0^{j*}+d_4^{j*})$ and $P_{f,j}^{(0)}=|d_0^{j*}+d_4^{j*}|^2/2$.
%
For the vacuum--Fock oscillator example of the main text, the measured observable is $\mathcal A=\hat a^\dagger\hat a$. The initial and postselection states are

\begin{align}
|\psi_i\rangle&=\frac{|0\rangle+|4\rangle}{\sqrt{2}},
\\
|\psi_f^1\rangle&=
\cos\frac{\theta_f}{2}|0\rangle+
e^{i\phi_f}\sin\frac{\theta_f}{2}
\left(
\cos\frac{\vartheta_f}{2}|1\rangle+
\sin\frac{\vartheta_f}{2}|4\rangle
\right).
\\
|\psi_f^2\rangle&=
\mathcal N_\chi(|\alpha\rangle+e^{i\chi}|-\alpha\rangle)
\end{align}
where
$\mathcal N_\chi=[2(1+e^{-2|\alpha|^2}\cos\chi)]^{-1/2}$,
with $\alpha$ the coherent-state amplitude and $\chi$ the relative phase.
For the first final state, Eq.~\eqref{add:osc-general-weak-value} gives
\begin{equation}
\mathcal A_w^1=
\frac{4e^{-i\phi_f}\sin(\theta_f/2)\sin(\vartheta_f/2)}
{\cos(\theta_f/2)+e^{-i\phi_f}\sin(\theta_f/2)\sin(\vartheta_f/2)}.
\label{add:osc-weak-one}
\end{equation}
For the even-cat postselection $\chi=0$, the corresponding analytic result is
\begin{equation}
\mathcal A_w^2=\frac{4(\alpha^*)^4}{\sqrt{24}+(\alpha^*)^4}.
\label{add:osc-weak-two}
\end{equation}
Only $n=0,4$ contribute to the overlaps, while $\mathcal N_\chi$ normalizes the full cat state. The diagonal instrument preserves the two-dimensional input support $\operatorname{span}\{|0\rangle,|4\rangle\}$. The components $d_n^j$ must retain the normalization of the full final state; renormalizing its projection onto this subspace would change the postselection and its success probability.
For either postselection, the subsequent record statistics use the corresponding $\mathcal A_w^j$ and $P_{f,j}^{(0)}$.

The number-operator measurement has binary outcomes, and all single-step Kraus operators are functions of $\mathcal A$. The weak-response result of Sec.~S2 therefore applies on the populated subspace. Since $\langle\mathcal A\rangle_i=2$, the leading displacement for either postselection is
\begin{equation}
\Delta\langle X\rangle_j\simeq\frac{\theta}{2}\left[\operatorname{Re}(\mathcal A_w^j)-2\right].
\end{equation}
Equation~\eqref{add:main-text-snr} gives the corresponding conditional ratio $\mathcal R_{\rm ps}^{\rm cond}\simeq\theta|\operatorname{Re}(\mathcal A_w^j)-2|\sqrt m$. At the captioned settings, $\mathcal A_w^1=-4(1+\sqrt2)$ and $\mathcal A_w^2=10$ are real and lie outside the interval $[0,4]$ spanned by the occupied number states; the first also lies below the lower bound of the full photon-number spectrum. For an unbounded number operator, the weak expansion must be controlled on the populated support, including the relevant higher moments and normalization. At finite strength, the record statistics are obtained from the exact operator and channel normalization below.

For finite measurement strength, write $C_n=\cos(n\theta/2)$ and $D_n=\sin(n\theta/2)$.
The exact complete-record operator is
\begin{equation}
K(q)=2^{-m/2}\sum_{n=0}^{\infty}
(C_n+D_n)^q(C_n-D_n)^{m-q}|n\rangle\langle n|.
\label{add:osc-exact-record}
\end{equation}
Equation~\eqref{add:exact-grouped-record} then also applies to the oscillator.
Its finite-sequence normalization is
\begin{equation}
\operatorname{Tr}[\Pi_f^j\Phi^m(\rho_i)]
=\frac12\sum_{n,l\in\{0,4\}}d_n^{j*}d_l^j
\left[\cos\!\left(\frac{(n-l)\theta}{2}\right)\right]^m.
\label{add:osc-exact-success}
\end{equation}
The exact distribution without postselection is $P_{\rm no}(q)=\binom mq\sum_{n\in\{0,4\}}|\langle n|K(q)|n\rangle|^2/2$, with mean $\langle X\rangle_{\rm no}=\sin(4\theta)/4$.
These results include accumulated measurement backaction and describe the finite-strength distributions in Fig.~\ref{SI-Fig.2} beyond the first-order weak-value approximation.
The retained sample size is determined by the same channel normalization, rather than by the initial overlap alone.

\bibliography{SI.ref}